\documentclass[]{aastex631}

\def\astrobjone{TIC 159102550} 
\def\astrobjtwo{V1068 Her} 
\def\astrobjthree{MW Pav} 
\def\astrobjfour{TIC 321576458} 

\begin{document}

\title{MULTI-COLOR AND TESS PHOTOMETRIC INVESTIGATION OF FOUR LOW MASS-RATIO CONTACT BINARY SYSTEMS}

\author[0000-0002-1230-4875]{Ahmed Waqas Zubairi}
\affiliation{Yunnan Observatories, Chinese Academy of Sciences (CAS), , 650216 Kunming, China}
\affiliation{University of Chinese Academy of Sciences, No.1 Yanqihu East Rd, Huairou District, Beijing, PR China, 101408}

\author{Zhao Ergang}
\affiliation{Yunnan Observatories, Chinese Academy of Sciences (CAS), , 650216 Kunming, China}
\affiliation{Center for Astronomical Mega-Science, Chinese Academy of Sciences, 20A Datun Road, Chaoyang District, Beijing, 100012, P. R. China}
\affiliation{Key Laboratory of the Structure and Evolution of Celestial Objects, Chinese Academy of Sciences, P. O. Box 110, 650216 Kunming, China}

\author{Qian Shengbang}
\affiliation{Department of Astronomy, Yunnan University, Kunming 650091, P. R. China}

\author{Zhou Xiao}
\affiliation{Yunnan Observatories, Chinese Academy of Sciences (CAS), , 650216 Kunming, China}
\affiliation{Center for Astronomical Mega-Science, Chinese Academy of Sciences, 20A Datun Road, Chaoyang District, Beijing, 100012, P. R. China}
\affiliation{Key Laboratory of the Structure and Evolution of Celestial Objects, Chinese Academy of Sciences, P. O. Box 110, 650216 Kunming, China}

\author{Eduardo Fernández Lajús}
\affiliation{Facultad de Ciencias Astronómicas y Geofísicas, Universidad Nacional de La Plata, Paseo del Bosques/n, 1900 La Plata, Buenos Aires, Argentina}
\affiliation{Instituto de Astrofísica de La Plata (CCT La Plata – CONICET/UNLP), 1900 La Plata, Argentina}

\begin{abstract}

We present the TESS and $BVR_cI_c$ light curves solution of four low mass-ratio contact binary systems {\astrobjone}), {\astrobjtwo}, {\astrobjthree} and {\astrobjfour}. Except {\astrobjthree}, all three systems have been studied for the first time. The $BVR_cI_c$ observations of southern contact binary system, {\astrobjthree}, is carried out using 60cm Helen Sawyer Hogg (HSH) telescope located at Argentina. The {\astrobjfour} is observed in $BVR_cI_c$ with 70cm Sino-Thai telescope at Lijiang station of Yunnan Observatories. The period analysis of {\astrobjone} show anti-correlation between primary and secondary minima and no long term variation is reported. The systems {\astrobjtwo} and {\astrobjthree} show increasing period at $dP/dt=1.247(9)\times 10^{-7} days/yr$ and $dP/dt=2.256(6)\times 10^{-7} days/yr$, respectively. Data for {\astrobjfour} is too few to determine any periodic variations. The light curve analysis using Wilson-Divinney model shows that systems are low mass-ratio contact binaries. Out of four targets, two systems {\astrobjone} and {\astrobjtwo}, are shallow contact binary systems with fill-out factor of 20\% and 14\%, respectively. The two contact binaries {\astrobjthree} and {\astrobjfour} are in deep contact state with fill-out factor 63\% and 61\%, respectively. {\astrobjtwo} shows EB-type light curve, however the temperature difference between the primary secondary component is only $17K$, which indicates that system is in thermal contact. To understand the evolutionary status of these systems, the components are plotted on the mass-luminosity diagram. The primary companions are in the ZAMS zone while the secondary components of all the systems are away from TAMS which indicates that secondary is more evolved than the primary components. {\astrobjtwo} and {\astrobjthree} are expected to evolve into a single rapidly rotating star provided that they meet the well-know Hut's criterion. Through statistical investigation of more than hundred low mass-ratio contact binary systems including our targets, we have found that all of the low mass-ratio contact binaries have undergone the mass-ratio inversion process. Based on our sample, the relationship between mass ratio and spin and orbital angular momentum ratio has been updated and proposed a new value of $q_{min}=0.0388$ for Darwin's stability.

\end{abstract}


\section{Introduction}
The two Roche lobe filled components of an eclipsing binary star formed a common convective envelope are referred as W UMa type eclipsing binary systems or simply contact binaries \citep{1956KopalEvolutionaryEBS, 1968aLucy}. Theses binaries are further divided into two categories; A-type and W-type \citep{1970BINNENDIJK, 1968aLucy}. The continuous variation in the light curve along with small difference between primary and secondary minima put them in EW-type eclipsing binary system. The small difference (nearly equal) depths in both minima indicates that the temperature of primary and secondary star is almost equal. However, the variation in magnitude at maxima shows the magnetic activity which can be explained with the presence of star spots \citep{1951Oconnel}. 
The orbital period study of contact binaries is also important. The orbital and spin angular momentum plays an important role in maintaining orbital stability \citep{1980Hut,1973Counselman} and provide useful information about the rate of mass transfer and angular momentum loss and help us to identify the potential merger candidates (e.g., \cite{2019Ferreira}). 

The widely accepted point of view about the formation of these system is via evolution from detached binaries due to AML or Kozai cycle. The strong interaction of binary companions makes their evolutionary scenario totally different compared to single star evolution. Some marginal contact systems are believed to be newly-formed contact binaries and are at the beginning of their contact phase \citep{2017QianLAMOSTEW,2018QianLAMOSTEA,2020RAA....20..163Q}. Their evolutionary process can be explained by the combination of angular momentum loss (AML) and thermal relaxation oscillation (TRO). Such system oscillate around a critical mass-ratio \citep{1976Lucy, 1976Flannery, 1977Robertson, 2001Qian,2003QianTRO,2014ZhuShallowCB}. These systems are expected to merge sooner into FK Comae type or blue straggler stars \citep{2005QianV857,2008QianEMPis,2014ZhuShallowCB, 2016Sriram,2016Zhuly,2022GAO}. Therefore, contact binaries, especially with low mass-ratio, work as natural laboratories for studying the mass and energy transfer, the angular momentum loss of the system, and stellar evolution \& merging processes. 


In this paper, we are presenting photometric investigation of four low mass-ratio contact binary systems. As a time of this writing, three systems (\astrobjone, \astrobjtwo and \astrobjfour) have not been studied in detailed before. We are presenting the first photometric investigation for these systems. The \astrobjthree (MW Pavonis) is a low-mass ratio contact binary system in southern hemisphere. The first photometric solution was presented by \cite{1968EggenMWPav}. Later on, \cite{1971WilliamonMWPav} and \cite{1977LapassetMWPav} also observed \astrobjthree but did not present any detailed solution. The radial velocity solution of \cite{2006RucinskiMWPav} reported the spectral type to be F3/IV-V along with spectroscopic mass ratio of $q_{sp}=0.228$ which was not consistent with their photometric mass-ratio ($q_{ph}=0.122$). \cite{2011SukantaMWPav}, uses the spectroscopic mass-ratio of \cite{2006RucinskiMWPav} and face problems in light curve fitting. However, they proposed the photometric mass-ratio of $q=0.200$ with $52\%$ degree of contact. More recently, \cite{2015AlvarezMWPav}, presented combined light and velocity curve solution. Their UBV photometric solution suggest a new mass-ratio $q=0.222$ and a high fill-out factor of $60\%$ for the system.

\section{Observations}
\subsection{TESS Observations}
The Transiting Exoplanet Survey Satellite (TESS) has rapidly increased the eclipsing binaries research \citep{2014TESS}. The obtained photometric data of a number of eclipsing binaries can be downloaded from Mikulski Archive for Space Telescopes (MAST) archive.\footnote{\url{https://mast.stsci.edu/portal/Mashup/Clients/Mast/Portal.html}}

The \astrobjone ($\alpha_{2000}=19^h18^m17.44^s, \, \delta_{2000}=+42^0 02'\,47.39''$) is listed as EW type binary in VSX catalog. The system is observed by TESS from May 14, 2020 to June 08, 2020 in sector 25 and 26 respectively. The data of \astrobjtwo (=TIC 193580427) ($\alpha_{2000}=17^h43^m23.08^s, \, \delta_{2000}=+47^0 51'\,41.79''$) is available from July 24, 2021 to August 20, 2021 in sector 41. 

Both of these two targets (\astrobjone\, and \astrobjtwo) were also observed by LAMOST \citep{2011Wu-spectral-library, 2012LAMOST}. The obtained atmospheric parameters are listed in Table \ref{tbl:atm-tic}. 

\begin{table}[h]
	\caption{Atmospheric parameter of {\astrobjone}\, and {\astrobjtwo}\, observed by LAMOST}
	\centering
	\begin{tabular}{lccccc}
		\hline
		Target  &  UT date & Spectral Class & $T_{eff}(K)$ & $log(g) (cm/s^2)$ & $[Fe/H] (dex)$  \\
		\hline
		{\astrobjtwo} & 18 May 2016   & F5 & $6605.09\pm12.23$ & $4.197\pm0.019$ &  $0.04\pm0.01$  \\
		{\astrobjone} & 08 March 2015 & G0 & $6158.85\pm50.28$ & $4.168\pm0.083$ &  $0.21\pm0.04$  \\
		
		\hline
	\end{tabular}
	\label{tbl:atm-tic}
	
\end{table}

\subsection{$BVR_cI_c$ Observations}

\astrobjthree\,(=TIC 372127422) ($\alpha_{2000}=20^h46^m27.74^s, \,  \delta_{2000}=-71^0 56'\,58.46''$, $V_mag = 8.79$) has been observed with 60cm Helen Sawyer Hogg (HSH) telescope at Complejo Astronomico El Leoncito (CASLEO), San Juan, Argentina. The new multi-color light curve is taken during three nights on August 09, 15, 20, 2019. These observations were carried out using SBIG STL1001E CCD camera with standard Jhonson's $BVR_cI_c$ filters attached to the HSH telescope giving $9.3'\times9.3'$ field of view. \astrobjthree is also observed by TESS from July 05, 2020 to September 30, 2020 in sector-27 using 2-minutes cadence. The differential magnitudes from HSH observations are determined using the standard aperture photometry procedure in Image Reduction and Data Analysis Facility (IRAF) software packages \citep{1986IRAF, 1993IRAF}. 

\astrobjfour ($\alpha_{2000}=15^h42^m12.72^s, \, \delta_{2000}=+37^0 52'\,46.50''$, $V_mag = 13.74$) has been observed with 70cm Sino-Thai telescope at Lijiang station of Yunnan Observatories. The first $BVR_cI_c$ light curve of \astrobjfour is obtained on April 13 and April 17, 2023. The telescope is equipped with Andor iKon 936 (2048x2048) CCD camera with standard Jhonson's $BVR_cI_c$ filters. The system  
is also observed by TESS from July 05, 2020 to September 30, 2020 in sector-27 using 2-minutes cadence. The differential magnitudes from the observations of 70cm Sino-Tahi telescope are determined using the standard aperture photometry procedure in Image Reduction and Data Analysis Facility (IRAF) software packages \citep{1986IRAF, 1993IRAF}.





\section{Orbital Period Investigation}

\subsection{Period Study of \astrobjone}

The period search of {\astrobjone}\, gives $0.488253 d$ from ROTSE All-Sky Surveys \citep{2000Akerlof}. No other database has any information on previously published period of {\astrobjone}. Therefore, the O-C values are determined by using epoch $MinI=BJD2459446.1702 + 0.488253E$.  

\begin{figure}[h]
	\centering
	\includegraphics[width=0.7\columnwidth]{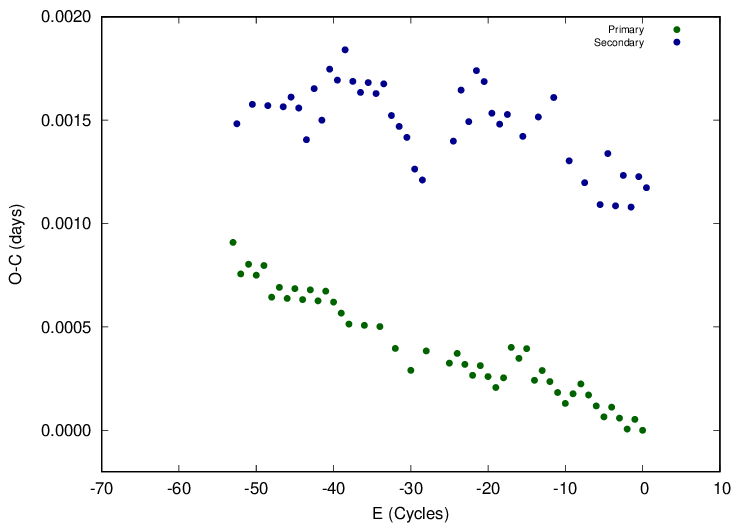}
	\caption{O-C diagram of {\astrobjone}. The primary and secondary minimas shows anti-correlation.}
	\label{fig:c1591-oc}
\end{figure}

\begin{figure}[h]
	\centering
	\includegraphics[width=0.49\columnwidth,height=6.5cm]{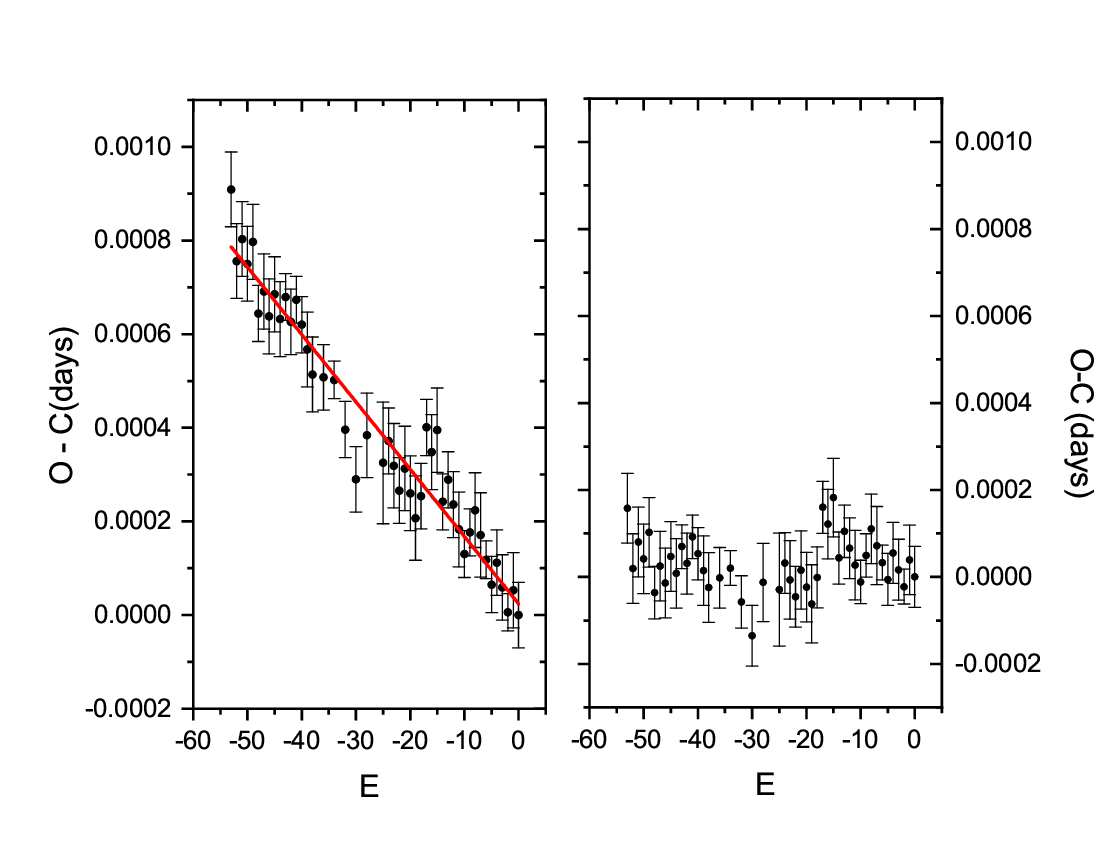} 
	\includegraphics[width=0.49\columnwidth,height=6.5cm]{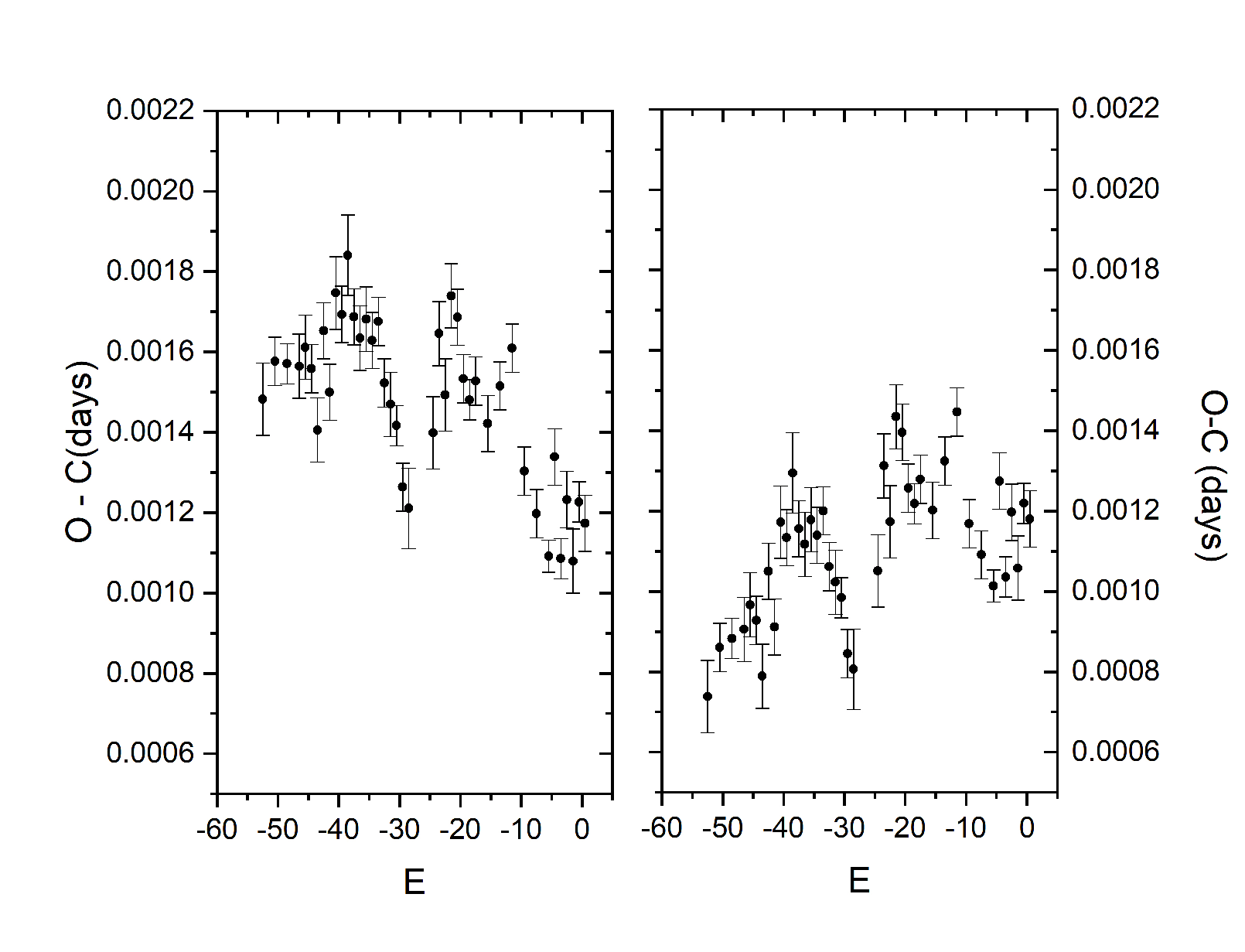}
	\caption{O-C diagram of {\astrobjone}. Left panel is calculated by the original period for the primary minima and display the revised one. The right panel is the same but for secondary minima.}
	\label{fig:c1591-oc2}
\end{figure}

The Figure \ref{fig:c1591-oc} shows that primary and secondary minimas are anti-correlated with each other. It further indicates that the period should be corrected from left upper panel of Figure \ref{fig:c1591-oc2} for the primary minima. The revised ephemeris and period are given in Equation \ref{eq:C1591-period}. A sample of times of primary and secondary minima are listed in Table \ref{tab:tom-c1591}.

\begin{equation}
\centering
MinI = BJD2459446.1702(7) + 0.48823883(8) \times E
\label{eq:C1591-period}
\end{equation}

We reconstruct the O-C diagram with the corrected period and ephemeris which is displayed in right panel of Figure \ref{fig:c1591-oc2}. The O-C for primary minima does not displays any large-scale variation as shown in upper panel of Figure \ref{fig:c1591-oc2}, which means that the revised period gives a better value than before. However, the corrected period from secondary minima shows a different behavior as shown in below panel of Figure \ref{fig:c1591-oc2}. The revised O-C values are also tabulated in Table \ref{tab:tom-c1591}. 

\begin{table}[h]
	\begin{center}
		\caption{Times of minima for {\astrobjone}. The BJD are +2400000.}
		\begin{tabular}{lccccccc} \hline
			%
			BJD  & Error  & Type & Method & E(Cycles) & $(O-C)_1$  & $(O-C)_2$ & Reference \\
			\hline
			
			59420.29370 & 0.00008 & p & TESS & -53.00 &0.00091 & 0.00016   & Present study \\
			59420.78180 & 0.00008 & p & TESS & -52.00 &0.00076 & 0.00002   & Present study\\
			59420.78180 & 0.00008 & p & TESS & -52.00 &0.00076 & 0.00002   & Present study\\
			59421.27010 & 0.00008 & p & TESS & -51.00 &0.00080 & 0.00008   & Present study\\
			........... & ....... & . & .... & ..... & .......  & .......  &    ......    \\
			59444.21730 & 0.00007 & p & TESS & -4.00  &0.00011 & 0.00006   & Present study\\
			59444.70550 & 0.00007 & p & TESS & -3.00  &0.00006 & 0.00002   & Present study\\
			59445.19370 & 0.00004 & p & TESS & -2.00  &0.00001 & -0.00002  & Present study\\
			59445.68200 & 0.00008 & p & TESS & -1.00  &0.00005 & 0.00004   & Present study\\
			........... & ....... & . & .... & ..... & .......  & .......  &    ......    \\
			59420.53840 & 0.00009 & s & TESS & -52.50 &0.00148  & 0.00074   & Present study\\
			59421.51500 & 0.00006 & s & TESS & -50.50 &0.00158  & 0.00086   & Present study\\
			59422.49150 & 0.00005 & s & TESS & -48.50 &0.00157  & 0.00088   & Present study\\
			59423.46800 & 0.00008 & s & TESS & -46.50 &0.00156  & 0.00091   & Present study\\
			........... & ....... & . & .... & ..... & .......  & .......  &    ......    \\
			59444.46240 & 0.00005 & s & TESS & -3.50  &0.00109  & 0.00104   & Present study\\
			59444.95080 & 0.00007 & s & TESS & -2.50  &0.00123  & 0.00120   & Present study\\
			59445.43890 & 0.00008 & s & TESS & -1.50  &0.00108  & 0.00106   & Present study\\
			59445.92730 & 0.00005 & s & TESS & -0.50  &0.00123  & 0.00122   & Present study\\
			........... & ....... & . & .... & ..... & .......  & .......  &    ......   \\
			\hline	
		\end{tabular}
	\end{center}
	\label{tab:tom-c1591}
\end{table}

\newpage

\subsection{Long term Period variation of \astrobjtwo}
\begin{figure}[h]
	\centering
	\includegraphics[width=0.7\columnwidth]{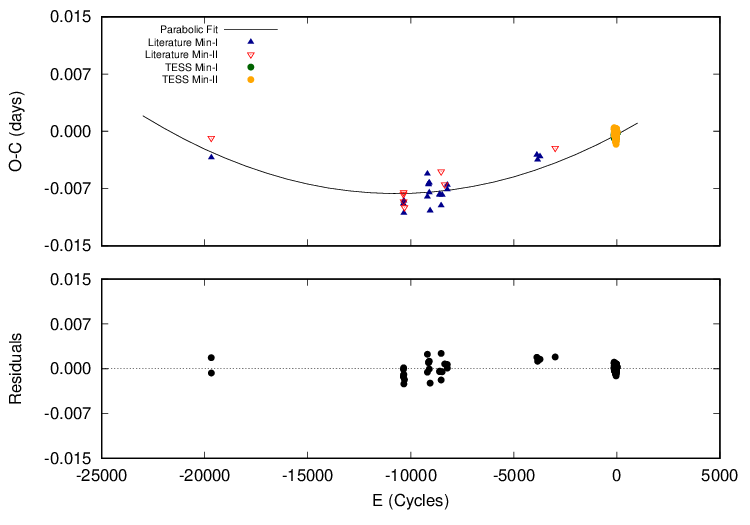}
	\caption{O-C diagram of {\astrobjtwo} with all available times of minima. The parabolic fit is shown in black line. Residuals are plotted in the bottom panel.}
	\label{fig:C1935-oc}
\end{figure}

The newly determined times of minima of {\astrobjtwo} along with values available in literature are listed in Table \ref{tab:timesofminimaC1935}. All HJDs are converted to BJD using \url{https://astroutils.astronomy.osu.edu/time/hjd2bjd.html}. The new O-C values are calculated by using the following epoch $MinI=BJD245934.9772+0.394304E$. The trend of O-C values indicate a upward parabolic fit, as shown in the Figure \ref{fig:C1935-oc}. The least square method yields that period of {\astrobjtwo}\, is increasing. We obtained the new ephemeris and period of {\astrobjtwo} as shown in Equation \ref{eq:C1935-period}.

\begin{equation}
\centering
MinI = 2459034.9767(6) + 0.39430544(2)\times E +6.73(5)\times 10^{-11} E^2
\label{eq:C1935-period}
\end{equation}

The quadratic term in Equation \ref{eq:C1935-period} leads us to determine the period increase at rate of $dP/dt=1.247(9)\times 10^{-7} days/yr$.

\begin{longtable}[h]{ccccccc}
	\caption{Times of minima for {\astrobjtwo}. The BJD are +2400000}\label{tab:timesofminimaC1935}\\
	\hline
	BJD  & Error  & Type & Method & E(Cycles) & O-C(days)  & Reference \\
	\hline
	\endhead
	
	\hline
	\endfoot
	
	51274.87645 &  - 	 &	s & ccd	 &-19680.50 &	-0.000881 &	\cite{2001Diethelm} \\
	51275.85965 &	-     &	p &	ccd  &-19678.00 &	-0.003441 &	\cite{2001Diethelm} \\
	54950.37256 &	-     &	p &	 R	 &-10359.00 &	-0.009505 &	\cite{2009OEJV..107....1B} \\
	54950.57096 &	-     &	s &	 R	 &-10358.50 &	-0.008257 &	\cite{2009OEJV..107....1B} \\
	54954.51306 &	-     &	s &	 I	 &-10348.50 &	-0.009197 &	\cite{2009OEJV..107....1B} \\
	54954.51426 &	-     &	s &	 R	 &-10348.50 &	-0.007997 &	\cite{2009OEJV..107....1B} \\
	54959.44036 &	-     &	p &	 I	 &-10336.00 &   -0.010697 & \cite{2011OEJV..137....1B} \\
	54959.44196 &	-     &	p &	 R	 &-10336.00 &	-0.009097 &	\cite{2011OEJV..137....1B} \\
	54971.46736 &	-     &	s &	-Ir	 &-10305.50 &	-0.009969 &	\cite{2010IBVS.5918....1H} \\
	55410.52626 &	-     &	p &	 I	 &-9192.00 &	-0.008568 &	\cite{2011OEJV..137....1B} \\
	55410.52926 &	-     &	p &	 R	 &-9192.00 &	-0.005568 &	\cite{2011OEJV..137....1B} \\
	55431.42596 &	-     &	p &	 V	 &-9139.00 &	-0.006980 &	\cite{2011OEJV..137....1B} \\
	55431.42606 &	-     &	p &	 B	 &-9139.00 &	-0.006880 &	\cite{2011OEJV..137....1B} \\
	55446.40846 &	-     &	p &	 I	 &-9101.00 &	-0.008031 &	\cite{2011OEJV..137....1B} \\
	55446.40977 &	-     &	p &	 R	 &-9101.00 &	-0.006731 &	\cite{2011OEJV..137....1B} \\
	55461.38967 &	-     &	p &	-Ir	 &-9063.00 &	-0.010383 &	\cite{2012IBVS.6010....1H} \\
	55645.53167 &	-     &	p &	 I	 &-8596.00 &	-0.008349 &	\cite{2013OEJV..160....1H} \\
	55645.53177 &	-     &	p &	 R	 &-8596.00 &	-0.008249 &	\cite{2013OEJV..160....1H} \\
	55672.34297 &	-     &	p &	-Ir	 &-8528.00 &	-0.009721 &	\cite{2012IBVS.6010....1H} \\
	55672.54457 &	-     &	s &	-Ir	 &-8527.50 &	-0.005273 &	\cite{2012IBVS.6010....1H} \\
	55692.45387 &	-     &	p &	-Ir	 &-8477.00 &	-0.008324 &	\cite{2012IBVS.6010....1H} \\
	55741.54607 &	-     &	s &	-Ir	 &-8352.50 &	-0.006972 &	\cite{2012IBVS.6010....1H} \\
	55791.42487 &	-     &	p &	 I	 &-8226.00 &	-0.007628 &	\cite{2013OEJV..160....1H} \\
	55791.42547 &	-     &	p &  R	 &-8226.00 &	-0.007028 &	\cite{2013OEJV..160....1H} \\
	57504.68028 &	-     &	p &	-I-U &-3881.00 &	-0.003093 &	\cite{2020BAVJ...33....1P}\\
	57518.48028 &	-     &	p &	-I	 &-3846.00 &	-0.003733 &	\cite{2017IBVS.6196....1H} \\
	57563.43138 &	-     & p &	-I-U &-3732.00 &	-0.003290 &	\cite{2020BAVJ...33....1P} \\
	57854.62599	&	-     & s & -I-U &-2993.50 &	-0.002185 &	\cite{2020BAVJ...33....1P}\\
	58983.71710 &	0.00011  &	p &	TESS &	-130.00 &	-0.000580 &	Present study \\
	58984.11170 &	0.00009	 &  p &	TESS &	-129.00 &	-0.000284 &	Present study \\
	58984.50590 &	0.00008 &	p &	TESS &	-128.00 &	-0.000388 &	Present study \\
	58984.90020 &	0.00007 &	p &	TESS &	-127.00 &	-0.000392 &	Present study \\
	........... & .......   &   . & .... & .....    & .......     & .......  \\
	59033.39960 &	0.00009 &	p &	TESS &	-4.00   &	-0.000384 &	Present study \\
	59033.79390 &	0.00007 &	p &	TESS &	-3.00   &	-0.000388 &	Present study \\
	59034.18820 &	0.00008 &	p &	TESS &	-2.00   &	-0.000392 &	Present study \\
	59034.58260 &	0.00008 &	p & TESS &	-1.00   &	-0.000296 &	Present study \\
	59034.97720 &	0.00008 &	p &	TESS &   0.00   &	 0.000000 &	Present study \\
	........... &   ....... &   . & .... & .....    & .......     & ....... \\
	58983.91530 &	0.00012 &	s &	TESS &	-129.50 &	 0.000468 &	Present study \\
	58984.30950 &	0.00008 &	s &	TESS &	-128.50 &	 0.000364 &	Present study \\
	58984.70370 &	0.00012 &	s &	TESS &	-127.50 &	 0.000260 &	Present study \\
	58985.49230 &	0.00011 &	s &	TESS &	-125.50 &	 0.000252 &	Present study \\
	58985.88650 &	0.00012 &	s &	TESS &	-124.50 &	 0.000148 &	Present study \\
	........... &   ....... &   . & .... & .....    & .......     & ....... \\
	59032.80840 &	0.00009 &	s &	TESS &	-5.50  &	-0.000128 &	Present study \\
	59033.20260 &	0.00009 &	s &	TESS &	-4.50  &	-0.000232 &	Present study \\
	59033.59710 &	0.00008 &	s &	TESS &	-3.50  &	-0.000036 &	Present study \\
	59034.38570 &	0.00007 &	s &	TESS &	-1.50  &	-0.000044 &	Present study \\
	59034.78000 &	0.00008 &	s &	TESS &	-0.50  &	-0.000048 &	Present study \\ 
	........... &   ....... &   . & .... & .....    & .......     & .......
 
	\label{tbl:timesofminimaC1935}
	
\end{longtable}

\subsection{Long term Period variation of \astrobjthree}

\begin{figure}[h]
	\centering
	\includegraphics[width=0.7\columnwidth]{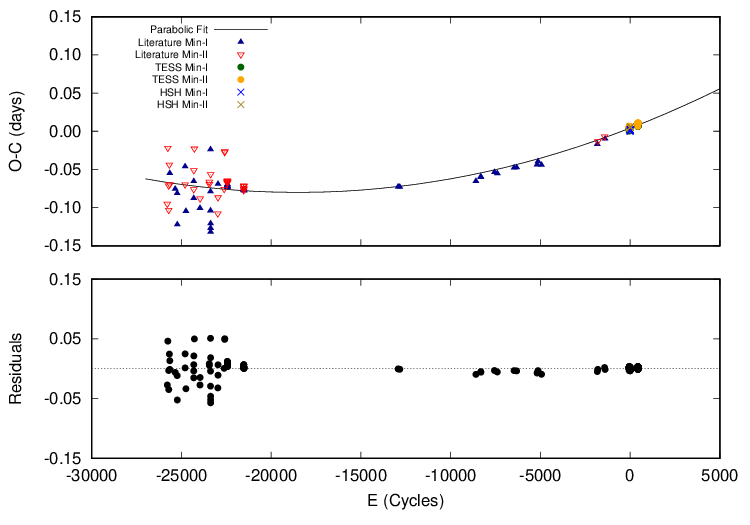} 
	\caption{The O-C diagram of {\astrobjthree} is constructed using $MinI=HJD2458704.6089+0.79498855E$. Black line shows parabolic fit. Residuals are plotted in the bottom panel.}
	\label{fig:mwpav-oc}
\end{figure}

The first observations were presented by \cite{1968EggenMWPav} with a period of 0.562979 days. Later on, the different period of MW Pav, was reported by many authors \citep{1971WilliamonMWPav, 1977LapassetMWPav}. The radial velocity solution of \cite{2006RucinskiMWPav} suggests that the period of system is $0.7949810 days$. The O-C gateway\footnote{\url{http://var2.astro.cz/ocgate}} and the spreadsheets of Bob Nelson, maintained at AAVSO database \citep{O-CNelson} contains most of the data which we have used during period investigation. Table \ref{tab:timesofminimamwpav} contains all the values that have been collected from the literature, $BVR_cI_c$ filters at HSH telescope and TESS data. All HJDs are converted into BJD. The O-C diagram indicates parabolic fit as shown in the Figure \ref{fig:mwpav-oc}. The least-square method was applied to obtain new ephemeris and period of {\astrobjthree} as shown in Equation \ref{eq:mwpav-period}.

\begin{equation}
\centering
MinI = 2458704.61313(8) + 0.79499753(5)\times E + 2.417(3)\times 10^{-10} E^2
\label{eq:mwpav-period}
\end{equation}

The quadratic term in Equation \ref{eq:mwpav-period} helped us to determine the period increase rate of $dP/dt= 2.220(9)\times 10^{-7} days/yr$. If we compare this value with \cite{2015AlvarezMWPav}, it can be noted that the rate of increase in period has been boosted up by $2.213\times 10^{-7}\, days/yr$.

\begin{longtable}{ccccccc}
	\caption{Times of minima for {\astrobjthree}. The BJD are +2400000}\label{tab:timesofminimamwpav}\\
	\hline 
	BJD  & Error  & Type & Method & E(Cycles) & O-C(days)  & Reference \\
	\hline
	\endhead
	
	\hline
	\endfoot

38204.5424  &   -      &  s  &  pg  &  -25786.5  &  -0.0950 &  \cite{1968EggenMWPav}  \\
38228.4654  &   -      &  s  &  pg  &  -25756.5  &  -0.0217 & \cite{1968EggenMWPav} \\ 
38263.3954  &   -      &  s  &  pg  &  -25712.5  &  -0.0711 &  \cite{1968EggenMWPav} \\
38267.3384  &   -      &  s  &  pg  &  -25707.5  &  -0.1031 &  \cite{1968EggenMWPav} \\
38295.2224  &   -      &  s  &  pg  &  -25672.5  &  -0.0437 &  \cite{1968EggenMWPav} \\
38314.2764  &   -      &  s  &  pg  &  -25648.5  &  -0.0694 &  \cite{1968EggenMWPav} \\
38316.2784  &   -      &  p  &  pg  &  -25646.0  &  -0.0549 &  \cite{1968EggenMWPav} \\
38555.5494  &   -      &  p  &  pg  &  -25345.0  &  -0.0754 &  \cite{1968EggenMWPav} \\
38641.4024  &   -      &  p  &  pg  &  -25237.0  &  -0.0812 &  \cite{1968EggenMWPav} \\
38649.3114  &   -      &  p  &  pg  &  -25227.0  &  -0.1221 &  \cite{1968EggenMWPav} \\
38992.4014  &   -      &  s  &  pg  &  -24795.5  &  -0.0696 &  \cite{1968EggenMWPav} \\
38994.4124  &   -      &  p  &  pg  &  -24793.0  &  -0.0461 &  \cite{1968EggenMWPav} \\
39029.3334  &   -      &  p  &  pg  &  -24749.0  &  -0.1046 &  \cite{1968EggenMWPav} \\
39374.3754  &   -      &  p  &  pg  &  -24315.0  &  -0.0876 &  \cite{1968EggenMWPav} \\
39376.3744  &   -      &  s  &  pg  &  -24312.5  &  -0.0761 &  \cite{1968EggenMWPav} \\
39378.3724  &   -      &  p  &  pg  &  -24310.0  &  -0.0655 &  \cite{1968EggenMWPav} \\
39380.3744  &   -      &  s  &  pg  &  -24307.5  &  -0.0510 &  \cite{1968EggenMWPav} \\
39404.2524  &   -      &  s  &  pg  &  -24277.5  &  -0.0227 &  \cite{1968EggenMWPav} \\
39654.1985  &   -      &  p  &  pg  &  -23963.0  &  -0.1006 &  \cite{1968EggenMWPav} \\
39656.1985  &   -      &  s  &  pg  &  -23960.5  &  -0.0880 &  \cite{1968EggenMWPav} \\
40064.0495  &   -      &  s  &  pg  &  -23447.5  &  -0.0661 &  \cite{1974LapassetIBVS} \\
40068.0215  &   -      &  s  &  pg  &  -23442.5  &  -0.0691 &  \cite{1974LapassetIBVS} \\
40120.0585  &   -      &  p  &  pg  &  -23377.0  &  -0.1038 &  \cite{1968EggenMWPav} \\
40120.0835  &   -      &  p  &  pg  &  -23377.0  &  -0.0788 &  \cite{1968EggenMWPav} \\
40120.9335  &   -      &  p  &  pg  &  -23376.0  &  -0.0238 &  \cite{1968EggenMWPav} \\
40122.8885  &   -      &  s  &  pg  &  -23373.5  &  -0.0563 &  \cite{1968EggenMWPav} \\
40124.0055  &   -      &  p  &  pg  &  -23372.0  &  -0.1318 &  \cite{1968EggenMWPav} \\
40124.0105  &   -      &  p  &  pg  &  -23372.0  &  -0.1268 &  \cite{1968EggenMWPav} \\
40124.0165  &   -      &  p  &  pg  &  -23372.0  &  -0.1208 &  \cite{1968EggenMWPav} \\
40440.0375  &   -      &  s  &  pg  &  -22974.5  &  -0.1077 &  \cite{1974LapassetIBVS} \\
40450.0135  &   -      &  p  &  pg  &  -22962.0  &  -0.0691 &  \cite{1974LapassetIBVS} \\
40451.9835  &   -      &  s  &  pg  &  -22959.5  &  -0.0866 &  \cite{1974LapassetIBVS} \\
40722.2905  &   -      &  s  &  pg  &  -22619.5  &  -0.0757 &  \cite{1974LapassetIBVS} \\
40746.1885  &   -      &  s  &  pg  &  -22589.5  &  -0.0273 &  \cite{1974LapassetIBVS} \\
40750.1645  &   -      &  s  &  pg  &  -22584.5  &  -0.0263 &  \cite{1974LapassetIBVS} \\
40862.6085  &   -      &  p  &  B   &  -22443.0  & -0.07318 &  \cite{1971WilliamonMWPav} \\
40862.6105  &   -      &  p  &  V   &  -22443.0  & -0.07118 &  \cite{1971WilliamonMWPav} \\
40862.6116  &   -      &  p  &  U   &  -22443.0  & -0.07008 &  \cite{1971WilliamonMWPav} \\
40863.8070  &   -      &  s  &  B   &  -22441.5  & -0.06716 &  \cite{1971WilliamonMWPav} \\
40863.8070  &   -      &  s  &  U   &  -22441.5  & -0.06716 &  \cite{1971WilliamonMWPav} \\
40863.8089  &   -      &  s  &  V   &  -22441.5  & -0.06526 &  \cite{1971WilliamonMWPav} \\
40864.6020  &   -      &  s  &  V   &  -22440.5  & -0.06715 &  \cite{1971WilliamonMWPav} \\
40864.6040  &   -      &  s  &  U   &  -22440.5  & -0.06515 &  \cite{1971WilliamonMWPav} \\
40864.6048  &   -      &  s  &  B   &  -22440.5  & -0.06435 &  \cite{1971WilliamonMWPav} \\
40870.5589  &   -      &  p  &  B   &  -22433.0  & -0.07266 &  \cite{1971WilliamonMWPav} \\
40870.5590  &   -      &  p  &  U   &  -22433.0  & -0.07256 &  \cite{1971WilliamonMWPav} \\
40870.5620  &   -      &  p  &  V   &  -22433.0  & -0.06956 &  \cite{1971WilliamonMWPav} \\
41587.6340  &   -      &  p  &  U   &  -21531.0  & -0.07722 &  \cite{1974LapassetIBVS} \\
41587.6352  &   -      &  p  &  B   &  -21531.0  & -0.07602 &  \cite{1974LapassetIBVS} \\
41587.6357  &   -      &  p  &  V   &  -21531.0  & -0.07552 &  \cite{1974LapassetIBVS} \\
41589.6266  &   -      &  s  &  U   &  -21528.5  & -0.07209 &  \cite{1974LapassetIBVS} \\
41589.6270  &   -      &  s  &  B   &  -21528.5  & -0.07169 &  \cite{1974LapassetIBVS} \\
41589.6275  &   -      &  s  &  V   &  -21528.5  & -0.07119 &  \cite{1974LapassetIBVS} \\
41592.8012  &   -      &  s  &  U   &  -21524.5  & -0.07744 &  \cite{1974LapassetIBVS} \\
41592.8036  &   -      &  s  &  V   &  -21524.5  & -0.07504 &  \cite{1974LapassetIBVS} \\
41592.8036  &   -      &  s  &  B   &  -21524.5  & -0.07504 &  \cite{1974LapassetIBVS} \\
41606.7131  &   -      &  p  &  U   &  -21507.0  & -0.07784 &  \cite{1974LapassetIBVS} \\
41606.7137  &   -      &  p  &  B   &  -21507.0  & -0.07724 &  \cite{1974LapassetIBVS} \\
41606.7149  &   -      &  p  &  V   &  -21507.0  & -0.07604 &  \cite{1974LapassetIBVS} \\
48426.1301	&	0.0026 &  p	 &  V   &  -12929.0	 & -0.07264 & \cite{1997HippCat} \\
48500.0637  &   -      &  p  &  V   &  -12836.0  &   -0.07297 &  \cite{1997HippCat} \\
51874.7978  &   -      &  p  &  V   &  -8591.0   &  -0.06529 &  \cite{2002PojmanskiASAS} \\
52085.4756	&  0.0020  &  p	 &  V   &  -8326.0   &  -0.05945 &  \cite{2002PojmanskiASAS}  \\
52096.6047	&  0.0033  &  p	 &  V   &  -8312.0	 &  -0.06019 &  \cite{2002PojmanskiASAS} \\
52694.4426  &  0.0017  &  p	 &  V   &  -7560.0   &  -0.05367 &  \cite{2002PojmanskiASAS} \\
52820.8443  &  0.0021  &  p	 &  V	&  -7401.0   &  -0.05515 &  \cite{2002PojmanskiASAS} \\
53564.9609  &  0.0023  &  p	 &  V	&  -6465.0   &  -0.04775 &  \cite{2002PojmanskiASAS} \\
53679.4396  &  0.0017  &  p	 &  V	&  -6321.0   &  -0.04741 &  \cite{2002PojmanskiASAS} \\
54584.9352  &  0.0018  &  p	 &  V	&  -5182.0   &  -0.04377 &  \cite{2002PojmanskiASAS} \\
54626.2789  &  0.0021  &  p	 &  V	&  -5130.0   &  -0.03947 &  \cite{2002PojmanskiASAS} \\
54782.8869  &  0.0021  &  p	 &  V	&  -4933.0   &  -0.04422 &  \cite{2002PojmanskiASAS} \\
57248.1738  &  0.003   &  p  &  V  &  -1832.0   &  -0.01686 &  \cite{2016Richards177} \\
57273.2198  &  0.003   &  s  &  V  &  -1800.5   &  -0.01300 &  \cite{2016Richards177} \\
57572.1418  &  0.004   &  s  &  V  &  -1424.5   &  -0.00670 &  \cite{2017Richards182} \\
57593.2058  &  0.004   &  p  &  V  &  -1398.0   &  -0.00989 &  \cite{2017Richards182} \\
58704.6097  &  0.0002  &  p  &  B  &  0.0  &  0.00000 &  Present study \\
58704.6097  &  0.0002  &  p  &  V  &  0.0  &  0.00000 &  Present study \\
58704.6108  &  0.0002  &  p  &  R  &  0.0  &  0.00110 &  Present study \\
58704.6116  &  0.0002  &  p  &  I  &  0.0  &  0.00190 &  Present study \\
58710.5781  &  0.0002  &  s  &  B  &  7.5  &  0.00600 &  Present study \\
58710.5779  &  0.0002  &  s  &  V  &  7.5  &  0.00582 &  Present study \\
58710.5775  &  0.0002  &  s  &  R  &  7.5  &  0.00539 &  Present study \\
58710.5770  &  0.0002  &  s  &  I  &  7.5  &  0.00493 &  Present study \\
58715.7412  &  0.0002  &  p  &  B  &  14.0  &  0.00166 &  Present study \\
58715.7421  &  0.0002  &  p  &  V  &  14.0  &  0.00256 &  Present study \\
58715.7414  &  0.0002  &  p  &  R  &  14.0  &  0.00186 &  Present study \\
58715.7413  &  0.0002  &  p  &  I  &  14.0  &  0.00176 &  Present study \\
58654.5281  &  0.0021  &  P  &  TESS  &  -63.0  &  0.00271 & Present study \\
58656.1180  &  0.0011  &  P  &  TESS  &  -61.0  &  0.00263 & Present study \\
58656.9098  &  0.0014  &  P  &  TESS  &  -60.0  &  -0.00056 & Present study \\
58660.0929  &  0.0004  &  P  &  TESS  &  -56.0  &  0.00259 & Present study \\
.........   &  ......  & .   &  .  &    ......  &   ....... &  .............\\
58662.4769  &  0.0008  &  P  &  TESS  &  -53.0  &  0.00162 & Present study \\
58663.2721  &  0.0007  &  P  &  TESS  &  -52.0  &  0.00183 & Present study \\
58665.6567  &  0.0007  &  P  &  TESS  &  -49.0  &  0.00147 & Present study \\
58666.4518  &  0.0006  &  P  &  TESS  &  -48.0  &  0.00158 & Present study \\
.........   &  ......  & .   &  .  &    ......  &   ....... &  .............\\
58654.1322  &  0.0006  &  S  &  TESS  &  -63.5  &  0.00429 & Present study \\
58656.5174  &  0.0016  &  S  &  TESS  &  -60.5  &  0.00453 & Present study \\
58658.9027  &  0.0008  &  S  &  TESS  &  -57.5  &  0.00489 & Present study \\
58659.6979  &  0.0002  &  S  &  TESS  &  -56.5  &  0.00510 & Present study \\
.........   &  ......  & .   &  .  &    ......  &   ....... &  .............\\
58660.4927  &  0.0016  &  S  &  TESS  &  -55.5  &  0.00489 & Present study \\
58662.0824  &  0.0020  &  S  &  TESS  &  -53.5  &  0.00465 & Present study \\
58662.8780  &  0.0005  &  S  &  TESS  &  -52.5  &  0.00528 & Present study \\
58663.6730  &  0.0004  &  S  &  TESS  &  -51.5  &  0.00528 & Present study \\
.........   &  ......  & .   &  .  &    ......  &   ....... &  .............
	\label{tbl:timesofminimamwpav}  
	
\end{longtable}

\subsection{Period Study of \astrobjfour}
\begin{figure}[h]
	\centering
	\includegraphics[width=0.7\columnwidth]{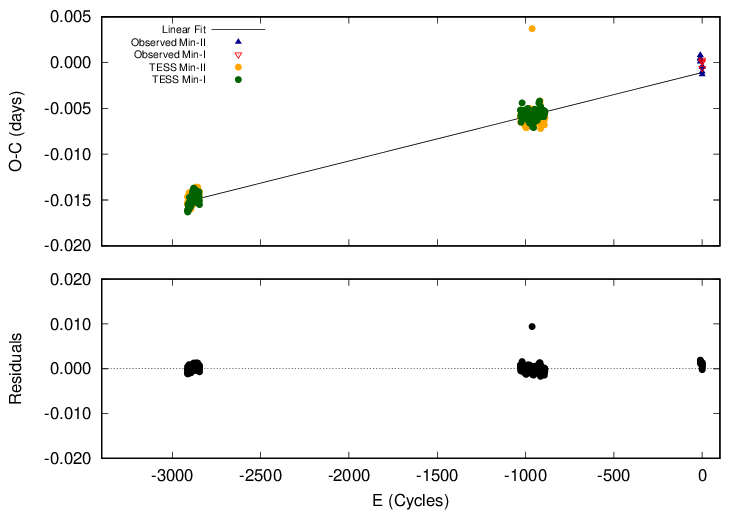} 
	\caption{The O-C diagram of {\astrobjfour} is constructed using $MinI=BJD2460052.39813+0.37614E$. Residuals are plotted in the bottom panel.}
	\label{fig:c154214-oc}
\end{figure}

The period of \astrobjfour was first reported to be 0.37614d by \citep{2014Drake-C154214}. No times of minimas are available for this target as of this writing. We have determine new times of primary and secondary minima from our $BVR_cI_c$ observations and TESS light curves using Kwee \& Wan Woerden algorithm \citep{1956Kewee}. The O-C diagram is shown in Figure \ref{fig:c154214-oc} and the times of primary and secondary minima are tabulated in Table \ref{tab:timesofminimac154214}. The figure clearly shows no variation in the orbital period. The new period and ephemeris are shown in Equation \ref{eq:c154214-period}.

\begin{equation}
\centering
MinI = 2460052.3995(2) + 0.376144829(1)\times E 
\label{eq:c154214-period}
\end{equation}

\begin{longtable}{ccccccc}
	
	\caption{A sample of Times of minima for {\astrobjfour}. The BJD are +2400000}\label{tab:timesofminimac154214}\\
	\hline 
	\footnotesize
	BJD  & Error  & Type & Method & E(Cycles) & O-C(days)  & Reference  \\
	\hline
	\endhead
	60052.39723  &  0.00034  &  P  &  B  &  0.0  &  -0.000903  &  Present study \\
	60052.39793  &  0.00023  &  P  &  Ic  &  0.0  &  -0.000203  &  Present study \\
	60052.39793  &  0.00020  &  P  &  Rc  &  0.0  &  -0.000203  &  Present study \\
	60052.39813  &  0.00019  &  P  &  V  &  0.0  &  -0.000003  &  Present study \\
	58956.30971  &  0.00049  &  P  &  TESS  &  -2914.0  &  -0.016460  &  Present study \\
	58956.68565  &  0.00064  &  P  &  TESS  &  -2913.0  &  -0.016660  &  Present study \\
	58957.06191  &  0.00311  &  P  &  TESS  &  -2912.0  &  -0.016540  &  Present study \\
	58957.43875  &  0.00115  &  P  &  TESS  &  -2911.0  &  -0.015840  &  Present study \\
	58957.81440  &  0.00109  &  P  &  TESS  &  -2910.0  &  -0.016330  &  Present study \\
	.........   &  ......  &  .   & ....  &   .....  &  ....... & .............\\
	59715.74690  &  0.00016  &  P  &  TESS  &  -895.0  &  -0.005930  &  Present study \\
	59716.12320  &  0.00023  &  P  &  TESS  &  -894.0  &  -0.005770  &  Present study \\
	59716.49920  &  0.00015  &  P  &  TESS  &  -893.0  &  -0.005910  &  Present study \\
	59716.87560  &  0.00026  &  P  &  TESS  &  -892.0  &  -0.005650  &  Present study \\
	59717.25170  &  0.00027  &  P  &  TESS  &  -891.0  &  -0.005690  &  Present study \\
	.........   &  ......  &  .   & ....  &   .....  &  ....... & .............\\
	60048.26083  &  0.00023  &  S  &  B  &  -11.0  &  0.000236  &  Present study \\
	60052.20843  &  0.00024  &  S  &  B  &  -0.5  &  -0.001633  &  Present study \\
	60052.20923  &  0.00026  &  S  &  $I_c$  &  -0.5  &  -0.000833  &  Present study \\
	60048.26043  &  0.00046  &  S  &  $I_c$  &  -11.0  &  -0.000164  &  Present study \\
	60048.26113  &  0.00019  &  S  &  $R_c$  &  -11.0  &  0.000536  &  Present study \\
	60052.20933  &  0.00028  &  S  &  $R_c$  &  -0.5  &  -0.000733  &  Present study \\
	60048.26073  &  0.00019  &  S  &  V  &  -11.0  &  0.000136  &  Present study \\
	60052.20873  &  0.00017  &  S  &  V  &  -0.5  &  -0.001333  &  Present study \\
	58956.12250  &  0.00051  &  S  &  TESS  &  -2914.5  &  -0.015600  &  Present study \\
	58956.49840  &  0.00498  &  S  &  TESS  &  -2913.5  &  -0.015840  &  Present study \\
	58956.87540  &  0.00253  &  S  &  TESS  &  -2912.5  &  -0.014980  &  Present study \\
	58957.25160  &  0.00112  &  S  &  TESS  &  -2911.5  &  -0.014920  &  Present study \\
	58957.62770  &  0.00158  &  S  &  TESS  &  -2910.5  &  -0.014960  &  Present study \\
	.........   &  ......  &  .   & ....  &   .....  &  ....... & .............\\	
	59715.93440  &  0.00023  &  S  &  TESS  &  -894.5  &  -0.006500  &  Present study \\
	59716.30990  &  0.00033  &  S  &  TESS  &  -893.5  &  -0.007140  &  Present study \\
	59716.68700  &  0.00022  &  S  &  TESS  &  -892.5  &  -0.006180  &  Present study \\
	59717.06300  &  0.00016  &  S  &  TESS  &  -891.5  &  -0.006320  &  Present study \\
	59717.43960  &  0.00020  &  S  &  TESS  &  -890.5  &  -0.005860  &  Present study \\
	
	\hline
\end{longtable}

\section{Light Curve Solutions}
We use Mode-3 of the Wilson-Divenney (WD) code to obtain the solution of light curves \citep{1971Wilson-Divinney,2014Wilson-VanHamme}. According to \cite{1967Lucygr1-gr2} and \cite{1969RucinskiA1-A2}, the gravity darkening coefficients ($g_1, g_2$) and albedos ($A_1, A_2$) should be $0.32$ and $0.5$, respectively as these values are suitable for binaries with convective envelopes. The limb darkening parameters were obtained from \cite{1993VanHamme}. As per Mode-3 constraint in the WD program, the fixed parameters were $g_1,\,g_2,\, A_1,\,A_2,\,x_1,\,x_2,\,y_1,\,y_2$. The gravitational potentials are kept equal ($\Omega_1 = \Omega_2$) for contact configuration.

\subsection{\astrobjone}
\begin{figure}[h]
	\centering
	\includegraphics[width=0.49\columnwidth]{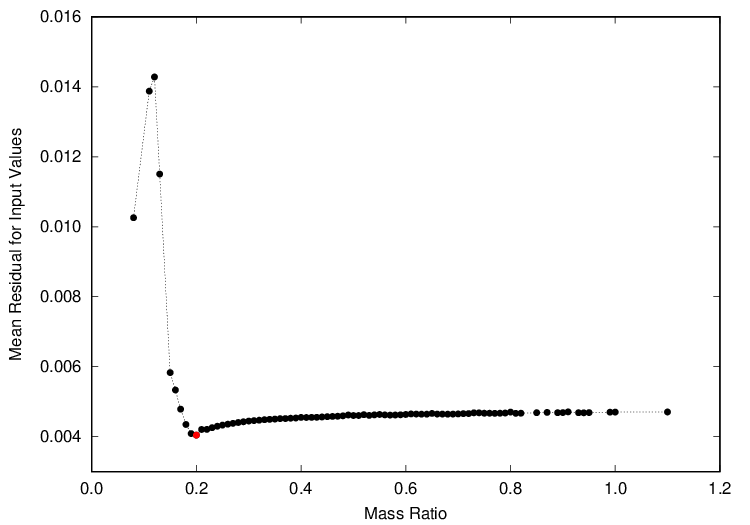}
    \includegraphics[width=0.49\columnwidth]{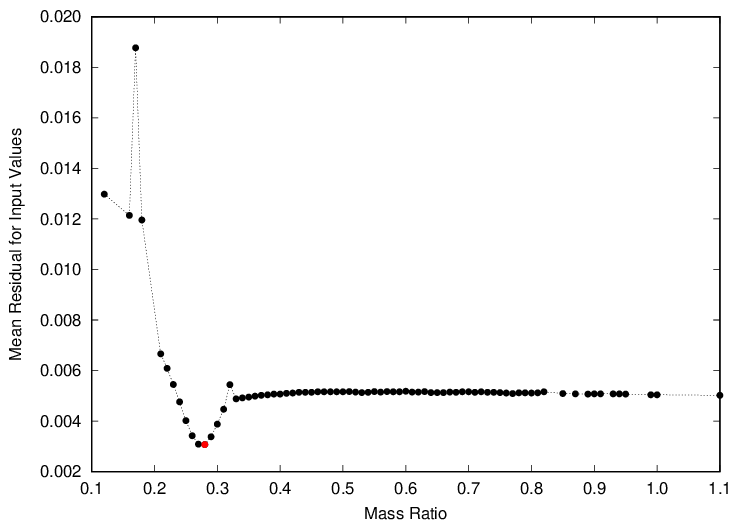}
    \includegraphics[width=0.55\columnwidth]{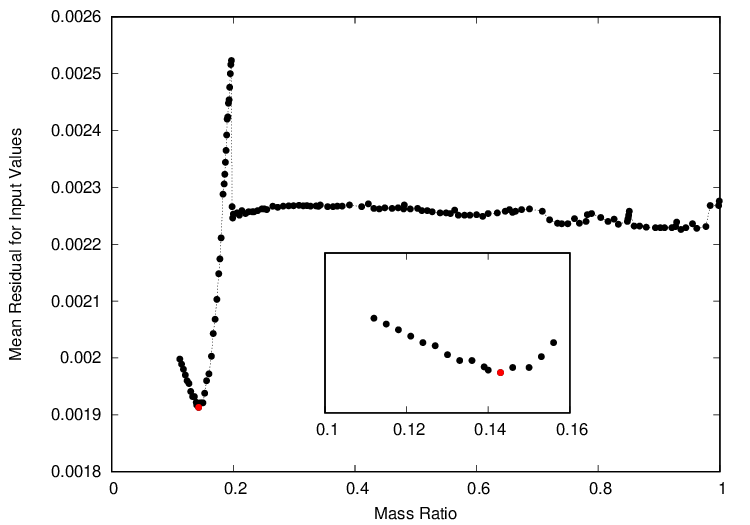}
	\caption{The q-search results of \astrobjone, (top left) \astrobjtwo (top right)\, and \astrobjfour (bottom).}
	\label{fig:gmr}
\end{figure}

In the absence of spectroscopic and photometric mass-ratio of this target, we apply the well-known q-search process to estimate the mass-ratio of the system. According to \cite{2003Pribulla} a reliable mass-ratio can be obtained for totally eclipsing binary systems. The results showed in Figure \ref{fig:gmr} indicates \ $q=0.20$ for \astrobjone. 

We use this q-value and keep it adjustable along with $T_2,\, \Omega_1,\, i,\, q,$ and  $L_1 $ to get the final convergent solution. The temperature of primary component ($T_1$) was opted from LAMOST data (Table \ref{tbl:atm-tic}). The modeled and observed data showed variation in amplitude of the light curve at phase 0.25 and 0.75. In order to cover this discrepancy in light curve solution, we introduced a cool spot on primary component for which parameters are listed in Table \ref{tbl:spotparameters}. The final light curve solution is shown in Figure \ref{fig:c1594-1935-lc}.

The obtained photometric parameters are listed in Table \ref{tab:c1591-1935-par}

\subsection{\astrobjtwo}
Since no photometric and spectroscopic mass ratio of this target was avaialbel in the literature, so we employed the q-search process to determine the unique value for the mass-ratio. The results of q-search method indicate the $q=0.28$ for \astrobjtwo. The results are displayed in Figure \ref{fig:gmr}.

\begin{figure}[h]
	\centering
	\includegraphics[width=0.49\columnwidth]{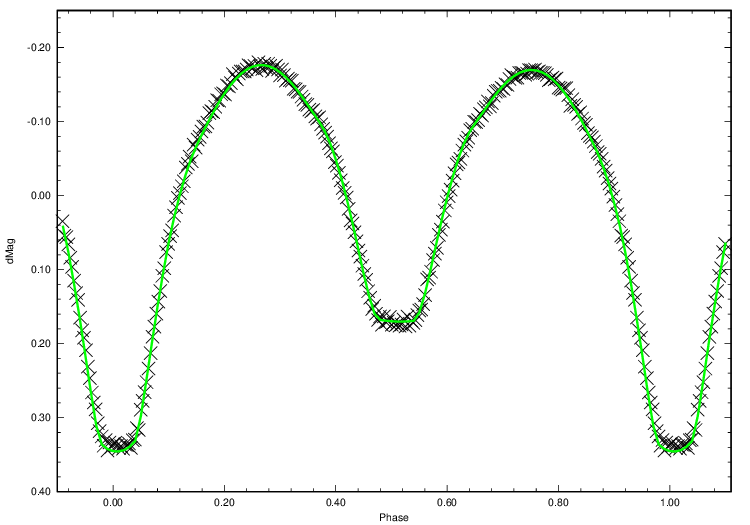}
     \includegraphics[width=0.49\columnwidth]{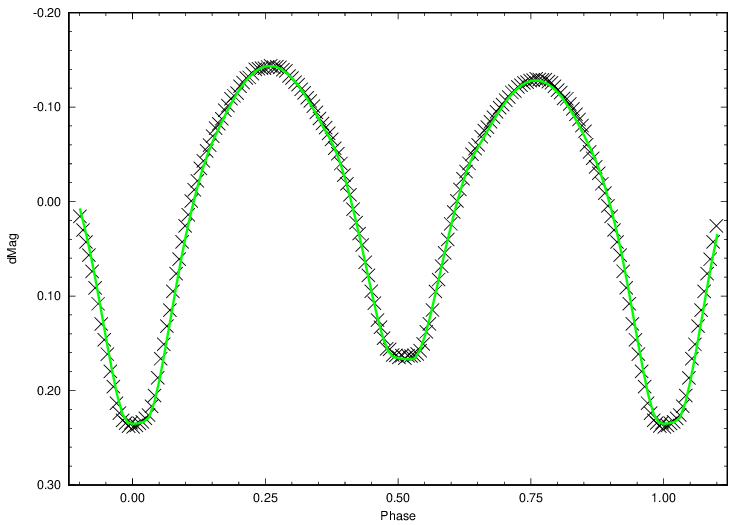}
	\caption{The light curve solution of \astrobjone\,(left) and \astrobjtwo (right). The modeled (solid green line) is superimposed over observed light curve.}
	\label{fig:c1594-1935-lc}
\end{figure}

We use this q-value and keep it adjustable along with $T_2,\, \Omega_1,\, i,\, q,$ and  $L_1 $ to get the final convergent solution. The temperature of primary component ($T_1$) was opted from Table \ref{tbl:atm-tic}. The convergent solution when superimposed on the observed data shows difference in Max-I and Max-II values. Therefore, we performed a spot solution on the light curve.  The obtained photometric parameters are listed in Table \ref{tab:c1591-1935-par} and the spot parameters are tabulated in Table \ref{tbl:spotparameters}. The final light curve solution is shown in Figure \ref{fig:c1594-1935-lc}.

\begin{table}
	\centering
	\caption{Photometric parameters of {\astrobjone\, and \astrobjtwo}}
	
	{	\begin{tabular}{lccccc}
			\hline
			
			& \multicolumn{2}{c}{\astrobjone} & & \multicolumn{2}{c}{\astrobjtwo}  \\
			\cline{2-3}
			\cline{5-6}
			Parameters & Without Spot & Spot && Without Spot & Spot\\
			\hline
			$q (m_2/m_1)$ & 0.207(4)  & 0.211(2) && 0.322(2)  & 0.266(3)   \\ 
			$i (deg)    $ & 77.14(4)  & 77.21(6) && 86.47(4)  & 83.53(3)  \\
			$\Omega_1 = \Omega_2$ & 2.220(4) & 2.229(8) && 2.458(8)  & 2.364(6)   \\
			$f (\%)$            & 18.88  & 20.74  &&  26.03 & 14.11  \\
			$T_1(K)\, (fixed) $ & 6109     & 6109  &&  6158     & 6158     \\
			$T_2(K)$     & 5977(16) & 6015(10) &&  5701(68) & 6143(57) \\
			$r_1(pole)$ & 0.492(2) & 0.490(1) && 0.461(1) & 0.470(1)  \\
			$r_1(side)$ & 0.537(3) & 0.535(1) && 0.498(1) & 0.509(1)  \\
			$r_1(back)$ & 0.562(3) & 0.560(1) && 0.528(1) & 0.535(1) \\
			$r_2(pole)$ & 0.242(9)  & 0.243(5)  && 0.280(3) & 0.260(4)   \\
			$r_2(side)$ & 0.253(11) & 0.254(6)  && 0.294(4) & 0.272(5)  \\
			$r_2(back)$ & 0.292(23) & 0.293(13) && 0.339(9) & 0.310(10)  \\
			$L_1/(L_1+L_2)$ & 0.844(4) & 0.842(2) && 0.825(2) & 0.824(6)  \\

			Mean Residual & 0.003706 & 0.001926  && 0.001666 & 0.000837  \\
	
			\hline
			
	\end{tabular}}
	\label{tab:c1591-1935-par}
\end{table}

\subsection{\astrobjthree}
The solution of \cite{2006RucinskiMWPav} suggests photometric mass-ratio of the system be $q_{ph}=0.122$ and $q_{sp}=0.228$. They also determined the spectral type of MW Pav to be F3/IV-V. \cite{2011SukantaMWPav}, uses the spectroscopic mass-ratio of \cite{2006RucinskiMWPav} and face problems in light curve fitting. However, they proposed the photometric mass-ratio of $q=0.200$ with a fill-out factor of $52\%$. More recently, \cite{2015AlvarezMWPav}, presented combined light and velocity curve solution. Their UBV photometric solution suggest a new mass-ratio $q=0.222$ and a high fill-out factor of $60\%$ for the system.

\begin{figure}[h]
	\centering
	
	\includegraphics[width=0.49\columnwidth]{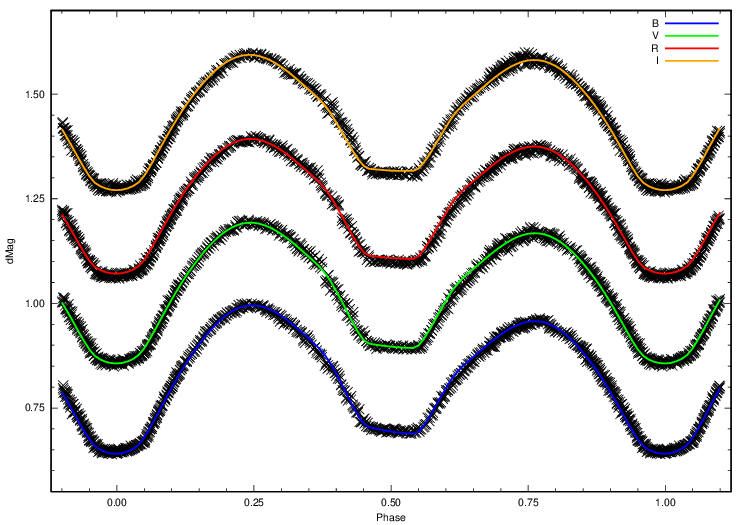} 
	\includegraphics[width=0.49\columnwidth]{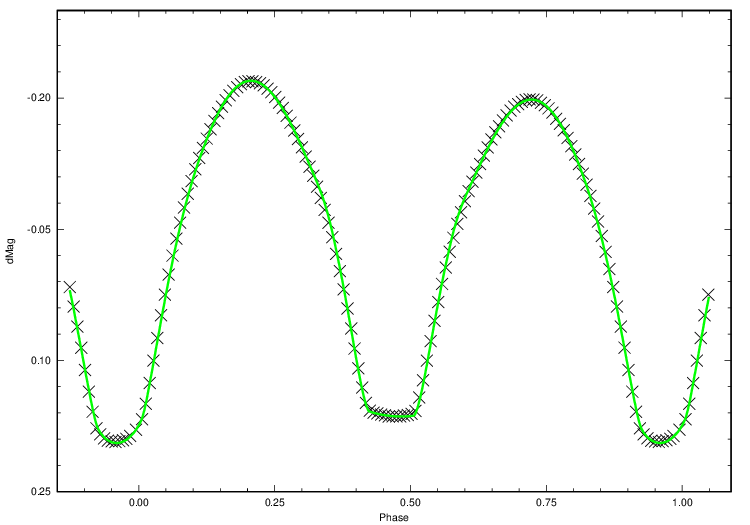} 
	
	\caption{The $BVR_cI_c$ light curve of  {\astrobjthree} is shown in the left and the TESS curve is on right.}
	\label{fig:mwpav-lc}
\end{figure}

We have analysed the new $BVR_cI_c$ and TESS light curves of this target. The initial value of mass ratio ($q=0.228$) and the proposed spectral type of {\astrobjthree} by \cite{2006RucinskiMWPav} help us to set the temperature of primary component at $T_1=6900K$ from \cite{2000Cox}. We use this q-value and get the convergent solution in Mode-3 of the WD program. The difference in Max-I and Max-II values compelled us to determine the spot parameters. We, therefore, run a spot solution for both, $BVR_cI_c$ and TESS, light curves. The final photometric solution is shown in Figure \ref{fig:mwpav-lc} and the parameters are listed in Table \ref{tab:mwpav-par}.

\begin{table}
	\centering
	\caption{Photometric parameters of {\astrobjthree}}
	
	{	\begin{tabular}{lccccc} 
			\hline
			
			& \multicolumn{2}{c}{HSH 60cm} & & \multicolumn{2}{c}{TESS}  \\
			\cline{2-3}
			\cline{5-6}
			
			Parameters & Without Spot & Spot && Without Spot & Spot\\
			\hline
			$q (m_2/m_1)$ & 0.200(3)  & 0.154(1) && 0.174(1) & 0.175(9) \\ 
			$i (deg)    $ & 84.28(98) & 80.20(40) && 82.17(66) & 81.79(24)  \\
			$\Omega_1 = \Omega_2$ & 2.167(6) & 2.045(3) && 2.099(13)  & 2.101(29)\\
			$f (\%)$            & 51.53 & 71.57  && 61.00 & 63.20 \\
			$T_1(K) (fixed)$ & 6900     & 6900  && 6900  & 6900 \\
			$T_2(K)$     & 6872(97) & 6987(57) && 6775(143) & 6838(12)\\
			$r_1(pole)$ &  0.502(1) & 0.523(1) && 0.513(2) & 0.514(5) \\
			$r_1(side)$ &  0.551(1) & 0.528(1) && 0.567(3) & 0.569(8) \\
			$r_1(back)$ &  0.579(1) & 0.609(1) && 0.594(3) & 0.596(8) \\
			$r_2(pole)$ & 0.252(6) & 0.237(3) && 0.245(11) & 0.244(2) \\
			$r_2(side)$ & 0.265(8) & 0.249(4) && 0.258(14) & 0.257(3) \\
			$r_2(back)$ & 0.319(21) & 0.314(15) && 0.319(42) & 0.317(10) \\
			$L_1/(L_1+L_2)_B$ & 0.827(4) & 0.806(4) && - & -  \\
			$L_1/(L_1+L_2)_V$ & 0.816(3) & 0.806(3) && - & - \\
			$L_1/(L_1+L_2)_{R_c}$ & 0.814(3) & 0.812(2) && - & -  \\
			$L_1/(L_1+L_2)_{I_c}$ & 0.820(2) & 0.822(2) && - & -  \\
			$L_1/(L_1+L_2)_{TESS}$ &  - & - && 0.815(1) & 0.808(8) \\
			Mean Residual & 0.001889 & 0.001401 && 0.002699 & 0.000843 \\
			
			\hline
			
	\end{tabular}}
	\label{tab:mwpav-par}
\end{table}

\begin{table}[h]
\begin{center}
		\caption{Star spot parameters for {\astrobjone}, {\astrobjtwo} and {\astrobjthree}}
		\begin{tabular}{lcccc} 
			\hline
			& Latitude(rad) & Longitude (rad) & Angular radius (rad) & Spot Temp. factor \\
			\hline
			
			{\astrobjone}   & 4.000$^{a}$ & 1.600$^{a}$ & 0.300$^{a}$ & 0.850$^{a}$ \\
			{\astrobjtwo}   & 0.140$^{a}$ & 0.172(19) & 1.196(55) & 0.899(6) \\
			{\astrobjthree} & 2.90$^{a}$ &  3.95(59)  & 0.60$^{a}$ & 1.38(14) \\
			{\astrobjthree}$_{TESS}$ & 0.667(20)  & 1.409(37)  & 0.223(33)  & 0.65$^{a}$  \\

			\hline
			\footnote  -Assumed
		\end{tabular}
		\label{tbl:spotparameters}
	\end{center}
\end{table}

\subsection{\astrobjfour}
\begin{figure}[h]
	\centering

	\includegraphics[width=0.49\columnwidth]{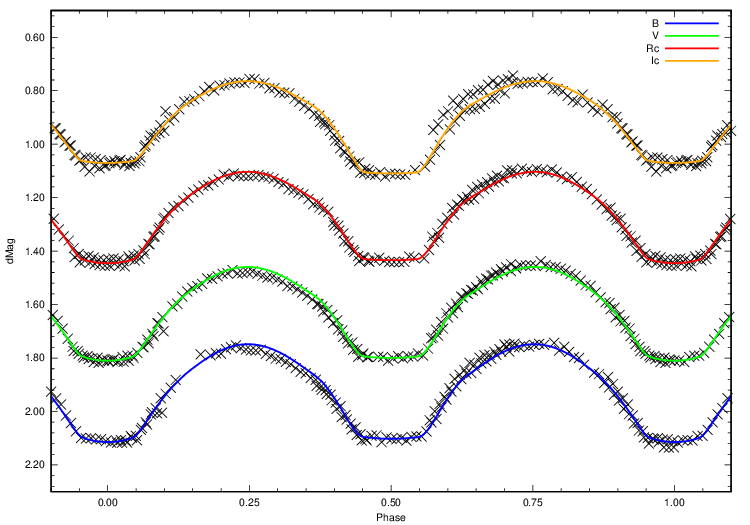} 
	\includegraphics[width=0.49\columnwidth]{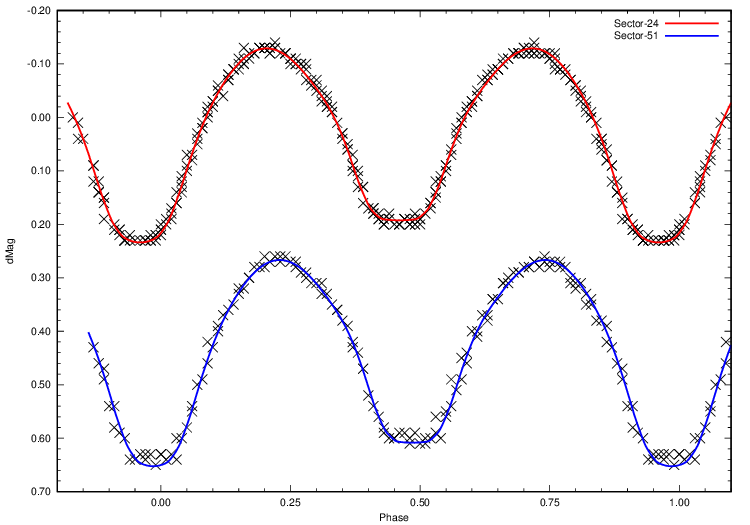} 
	
	\caption{Combined light curves of \astrobjfour. The $BVR_cI_c$ light curve of is shown in the left and the TESS curves are on right side. Solid line represents the modeled solution.}
	\label{fig:c154214-lc}
\end{figure}

Since this target is also observed for the first time and no other photometric and spectroscopic mass-ratio is available in the literature. Therefore, we employed the q-search process to determine the initial mass-ratio of the system. The results are shown in Figure \ref{fig:gmr} which indicates the q=0.143. By using this q-value and keep it adjustable along with $T_2,\, \Omega_1,\, i,\, q,$ and  $L_1 $ to get the final convergent solution. The temperature of primary component ($T_1$) was taken from GAIA data. The modeled light curve fits well over the observed light curve and therefore not spot solution was performed. The fitted light curves are shown in Figure \ref{fig:c154214-lc}.

\begin{table}[h]
	\centering
	\caption{Photometric parameters of {\astrobjthree}}
	{	\begin{tabular}{lccccc}
			\hline
			
			& \multicolumn{2}{l}{Sino-Thai} & & \multicolumn{2}{l}{TESS}  \\
			\cline{2-2}
			\cline{4-6}
			
			Parameters & 70cm telescope && Sector-24 && Sector-51\\
			\hline
		$q (m_2/m_1)$ & 0.153(1) && 0.147(1) && 0.167(2) \\ 
		$i (deg)    $ & 82.57(6) && 89.69(4) && 88.40(1) \\ 
		$\Omega_1 = \Omega_2$ & 2.071(4) && 2.030(5) && 2.082(6)  \\
		$f(\%)$  & 41.7 && 68.6 && 61.2 \\
		$T_1(K)(fixed)$  & 6140 && 6140 && 6140 \\
		$T_2(K)$  & 6490(116) && 6001(13) && 5982(25)  \\
		$r_1(pole)$ & 0.516(1) && 0.526(1)  && 0.517(1)  \\
		$r_1(side)$ & 0.571(1) && 0.585(1)  && 0.573(1)  \\
		$r_1(back)$ & 0.595(1) && 0.611(1)  && 0.599(2)  \\
		$r_2(pole)$ & 0.228(3)  && 0.234(5)  && 0.239(6)  \\
		$r_2(side)$ & 0.238(4)  && 0.246(7)  && 0.252(7)  \\
		$r_2(back)$ & 0.285(10) && 0.310(24) && 0.309(21) \\
		$L_1/(L_1+L_2)_B$     & 0.839(1) && - && -   \\
		$L_1/(L_1+L_2)_V$     & 0.836(7) && - && -   \\
		$L_1/(L_1+L_2)_{R_c}$ & 0.836(5) && - && -   \\
		$L_1/(L_1+L_2)_{I_c}$ & 0.815(5) && - && -   \\					
		$L_1/(L_1+L_2)_{TESS}$ & -    && 0.844(1) && 0.833(3) \\
		Mean Residual &  0.001179 && 0.001896 && 0.003015 \\

\hline

			\hline
			
	\end{tabular}}
	\label{tab:c154214-par}
\end{table}

\section{Discussion and Conclusion}

\subsection{\astrobjone}
By combining all times of observed primary and secondary minima, we have constructed the O-C diagrams to investigate the period changes. The O-C diagram of {\astrobjone} hints to correct the period. After period correction, we determined new values of O-C and reconstructed the diagram. The O-C for primary minima displays no large-scale change as shown in upper panel of Figure \ref{fig:c1591-oc2}, which means that the revised period gives a better value than before. If we apply the same to secondary minima, it shows a different behavior as shown in below panel of Figure \ref{fig:c1591-oc2}. The O-C of many targets has variable change, which include regular and irregular, such as long-term increasing and decreasing \citep{2021zhao,2022PASJLi, 2022zhao}. This can be explained by mass-transfer between both components, a third body, magnetic activities or apsidal motion \citep{2019ApJZhao, 2022ApJLi}. Sudden change and even anti-correlations between the primary and secondary eclipse timing has also been reported. This anti-correlation behaviour can be explained with rapid variation in light curve due to migration of star spot over time. This phenomena can deviate the eclipse times from theoretical phase which results in different trend in O-C diagram over a small timescale \citep{2013ApJtran, 2019MNRASzhou, 2023Chang-HHUMa}. 

In the absence of spectroscopic parameters, absolute values of binary components cannot be determined directly. However, if we assume that the primary component is a main-sequence star, then we can estimate the mass of secondary component from \cite{2000Cox}. The absolute parameters are tabulated in Table \ref{tbl:absparameters}. The photometric solution suggest that system is an A-type low mass-ratio ($q=0.211(2)$) contact binary with a weak degree of contact ($20.74\%$). It's secondary component is an evolved star while the more massive primary is still on the main-sequence as shown in Figure \ref{fig:ML-dia}. We suggest continuous monitoring of this target.

\subsection{\astrobjtwo}
The period investigation of {\astrobjtwo} suggest that binary is going through increase in period by $dP/dt=1.247(9)\times 10^{-7} days/yr$. The continuous increase in period indicates that binary is undergoing a process of mass transfer from less massive secondary to more massive primary component. We can use the well-known equation \citep{1986SinghML, 2017Carroll}

\begin{equation}
\frac{dP/dt}{P} = 3 \, \frac{dM_2}{dt} \, \left(  \frac{1}{M_2} - \frac{1}{M_1} \right) 
\label{eq:mass-transfer}
\end{equation}

and determine the mass transfer rate of $4.304(5)\times10^{-8}M_{\odot} yr^{-1}$ at thermal time-scale of $6.481(5) \times 10^6 \, yr$ for {\astrobjtwo}.

Since no absolute parameters of this target was previously published, so we have estimated the mass of primary component from \cite{2000Cox}. All the parameters are listed in Table \ref{tbl:absparameters}. The photometric solution indicates the mass-ratio of $q=0.266(3)$ and shallow degree of contact ($f=14.11\%$). The system lies in A-subtype category of contact binaries. Its more massive primary component is a main-sequence star while the less massive secondary companion is an evolved star. With the transfer of mass from secondary to primary component, the mass-ratio of the system will further decrease. If the mass transfer process is conservative, the period change of \astrobjtwo suggest that the system will oscillate around a critical mass-ratio. Since both component of binary are in thermal contact ($\Delta T = 15K$), they may not be oscillating between contact and semi-detached state. When the mass-ratio decreases to critical value the period will decrease which indicates the reversal in mass transfer direction. Therefore, the mass-ratio will start increasing and as soon as it increases the critical value, the period of \astrobjtwo\, will start increasing again and the system will keep on oscillating. A critical AML rate will keep the \astrobjtwo in shallow contact stage. The larger AML rate will make the stars to reach at the outer Lagrangian point ($L_2$) before merger into a single star \citep{1981Rahunen,2003QianTRO,2013QianUZUMi}.


\subsection{\astrobjthree}
The long term period analysis of {\astrobjthree} indicates the increase in period by $2.220(9)\times 10^{-7}\, days/yr$. The continuous increase in period indicates that both systems are undergoing a process of mass transfer from less massive secondary to more massive primary component. We have used the Equation \ref{eq:mass-transfer} and determined that system is transferring the mass from less massive secondary to more massive primary component at the rate of $2.307(6)\times10^{-8}M_{\odot} yr^{-1}$ in thermal time-scale of $1.11(4) \times 10^7 \, yr$.

The $BVR_cI_c$ light curve analysis shows that system is a W-subtype low mass-ratio ($q=0.154(1)$) contact binary in deep contact stage ($f=71.57\%$). However, light curve solution of TESS data put it in A-subtype category. It is possible that binary has now changed from W-type to A-type contact system. Such phenomena has also been reported by \cite{2005QianFGHya} and \cite{2008QianEMPis}. 
We have opted the radial velocity data from \cite{2006RucinskiMWPav} and combined it with our photometric solution to determine the absolute parameters of {\astrobjthree}. They are presented in Table \ref{tbl:absparameters}. With the increase in period and the transfer of mass from secondary to primary component, the mass-ratio of the system will further decrease. The increasing period will widen the orbit, however, the strong degree of contact, ($71.51\%$) for HSH observations and ($63.20\%$) for TESS, will hold the binary in contact configuration for a longer duration of time until the Hut's criteria \citep{1980Hut} is achieved ($J_{orbital} \le 3J_{spin}$). The {\astrobjthree} will eventually evolve into single rapidly rotating star.

\subsection{\astrobjfour}
The O-C diagram of this target (Figure \ref{fig:c154214-oc}) does not hint of any period variation during the short time span, therefore, period change and mass transfer rate cannot be determined. Assuming that the primary component is a main-sequence star, we have estimated the absolute parameter from \cite{2000Cox} and shown in Table \ref{tbl:absparameters}. The photometric solution suggest that system is an W-type low mass-ratio ($q=0.153(1)$) with moderate fill-out factor of  $41.7\%$. However, TESS light curve analysis put it in A-type category with higher contact degree of $61.2\%$. It's secondary component is an evolved star while the more massive primary is still on the main-sequence as shown in Figure \ref{fig:ML-dia}. More times of primary and secondary minima will help to study the orbital period in detail. We suggest to monitor the target for a longer duration.

\begin{table}
	\begin{center}
		\caption{Absolute parameters four contact binaries}
		\resizebox{\textwidth}{!}
		{	\begin{tabular}{lccccccc}
				\hline
				System & $M_1$ & $M_2$ &  $R_1$  & $R_2$ & $L_1$ &  $L_2$ & a \\
				\hline
				{\astrobjone}    & $1.400M_{\odot}$ & $0.295M_{\odot}$ & $1.633R_{\odot}$ & $0.817R_{\odot}$ & $4.535L_{\odot}$ & $0.783L_{\odot}$ & $3.104R_{\odot}$\\
				{\astrobjtwo}    & $1.050M_{\odot}$ & $0.279M_{\odot}$ & $1.252R_{\odot}$ & $0.696R_{\odot}$ & $2.022L_{\odot}$ & $0.618L_{\odot}$ & $2.482R_{\odot}$\\
				{\astrobjthree}  & $1.670M_{\odot}$ & $0.257M_{\odot}$ & $2.481R_{\odot}$ & $1.193R_{\odot}$ & $12.518L_{\odot}$ & $3.043L_{\odot}$ & $4.485R_{\odot}$\\
				{\astrobjfour}    & $1.200M_{\odot}$ & $0.183M_{\odot}$ & $1.417R_{\odot}$ & $0.632R_{\odot}$ & $2.561L_{\odot}$ & $0.637L_{\odot}$ & $2.527R_{\odot}$\\
				\hline
				
		\end{tabular}}
		
		\label{tbl:absparameters}
	\end{center}
\end{table}


\subsection{Statistics on some low mass ratio binaries}
During recent years, many authors have paid much attention to the statistical studies on contact binary systems. The analysis of orbital period distribution of several authors (e.g., \citep{1976Lucy}, \citep{2007MNRAS.382..393R}) showed a strong and sharp peak at nearly 0.22 days which is considered as short-period limit for contact binaries. \cite{2017QianLAMOSTEW}, through investigation of more than 40,000 systems reported 0.29 days and \cite{2020ZhuLiKIC} established 0.30 days while analyzing 446 systems. Not only this, \cite{1967MmRAS..70..111E}, established an important period-color relationship for contact binaries which helped to the model of thermal relaxation oscillation theory. \cite{2015Yang} also performed statistical analysis on the evolutionary scenarios of contact binary systems but his analysis was limited to deep contact binaries. We have discussed the distribution of low mass-ratio contact binary systems on the Mass-Luminosity (ML) diagram to study their evolutionary track. We have also calculated their initial mass, age and momentum ratio of these system. We have collected absolute parameters of 122 contact binaries, including our present study, to understand the evolutionary scenario of our target binaries. The data contains 47 extreme low mass-ratio (ELMR) binaries with $q\leq0.100$ tabulated in Table \ref{tab:abselmr} and 75 low mass-ratio (LMR) binaries with $q=0.10-0.25$ listed in Table \ref{tab:abslmr}.

\newpage
\begin{table}[h]
	
	\caption{Absolute parameters of extreme low mass-ratio contact binaries} 	
	
	\resizebox{\textwidth}{!}
	{
		\begin{tabular}{lccccccccccccccl}
			
			\label{tab:abselmr}\\
			\hline
			Star  &  $q$    &  $P$  &   $dP/dt$   &  $f $  &  $i$    &  $T_1$ &  $T_2$ &  $M_1 $  &  $M_2$  &  $R_1$  &   $R_2$   &   $L_1$  &  $L_2$  &  $a$  & Ref.\\
			
			&  ($m_2/m_1$)  & (d) & ($\times10^{-7}$ d)  & (\%)  & $(^0)$ &  (K)  & (K)  & ($M_{\odot}$) &  ($M_{\odot}$) & ($R_{\odot}$)  &  ($R_{\odot}$) & ($L_{\odot}$) & ($L_{\odot}$) & ($R_{\odot}$) & \\
			
			\hline

			V1187 Her  &  0.044  &  0.310760  &  -1.5  &  80.00  &  65.90  &  6250  &  6651  &  1.200  &  0.050  &  1.370  &  0.380  &  2.570  &  0.254  &  2.076  &  \cite{2019Caton} \\
			
			TYC 4002-2628-1  &  0.048  &  0.367058  &  0.0162  &  5.00  &  69.90  &  6032  &  6044  &  1.110  &  0.054  &  1.450  &  0.390  &  2.497  &  0.182  &  2.265 & \cite{2023GuoTYC}\\
			
			VSX J082700.8+462850  &  0.055  &  0.277158  &  -9.52  &  19.00  &  68.70  &  5870  &  5828  &  1.060  &  0.060  &  1.150  &  0.320  &  1.409  &  0.106  &  1.854 & \cite{2021Li} \\
			
			IP Lyn  &  0.055  &  0.489115  &  3.82  &  21.40  &  75.54  &  6677  &  6410  &  1.903  &  0.105  &  2.070  &  0.586  &  7.641  &  0.520  &  3.289 & \cite{2023IPLyn}\\
			
			KIC 4244929  &  0.059  &  0.341403  &  -  &  81.00  &  70.60  &  5857  &  5867  &  1.481  &  0.087  &  1.521  &  0.477  &  2.443  &  0.242  &  2.384 & \cite{2016SenavcKIC}\\
			
			KIC 9151972  &  0.059  &  0.386796  &  -  &  76.00  &  70.10  &  6040  &  5982  &  1.606  &  0.095  &  1.696  &  0.528  &  3.435  &  0.320  &  2.662  & \cite{2016SenavcKIC}\\
			
			ASAS J083241+2332.4  &  0.065  &  0.311300  &  8.85  &  50.65  &  82.74  &  6300  &  6667  &  1.220  &  0.080  &  1.340  &  0.420  &  2.538  &  0.313  &  2.106  & \cite{2016Sriram}\\
			
			V857 Her  &  0.065  &  0.382231  &  2.9  &  83.80  &  85.43  &  8300  &  8513  &  2.180  &  0.142  &  1.854  &  0.616  &  14.636  &  1.788  &  2.929 & \cite{2005QianV857} \\
			
			KIC 8539720  &  0.067  &  0.744499  &  -  &  47.00  &  71.10  &  6351  &  5972  &  2.438  &  0.163  &  2.955  &  0.929  &  12.746  &  0.985  &  4.745 & \cite{2016SenavcKIC}\\
			
			CW Lyn  &  0.067  &  0.812401  &  -  &  -  &  77.50  &  6532  &  6284  &  1.680  &  0.110  &  2.810  &  0.960  &  12.897  &  1.289  &  4.440& \cite{2004CWLyn}\\
			
			ASAS J104422-0711.2  &  0.073  &  0.611711  &  -  &  83.00  &  72.70  &  7200  &  7190  &  1.480  &  0.110  &  2.210  &  0.760  &  11.776  &  1.385  &  3.533 & \cite{2023WadhwaASAS}\\
			
			KIC 3127873  &  0.073  &  0.671526  &  -  &  65.00  &  80.50  &  6069  &  5791  &  2.268  &  0.166  &  2.690  &  0.899  &  8.808  &  0.815  &  4.333 & \cite{2016SenavcKIC}\\
			
			KIC 12352712  &  0.073  &  0.722065  &  -  &  57.00  &  82.50  &  6667  &  6399  &  2.377  &  0.174  &  2.859  &  0.944  &  14.489  &  1.341  &  4.619 & \cite{2016SenavcKIC}\\
			
			NSVS 2569022  &  0.077  &  0.287797  &  -  &  1.40  &  76.29  &  6100  &  6100  &  1.170  &  0.090  &  1.190  &  0.380  &  1.759  &  0.179  &  1.977 & \cite{2018NSVS2569022}\\
			
			ZZ PsA  &  0.078  &  0.374050  &  -  &  97.00  &  75.25  &  6514  &  6703  &  1.213  &  0.095  &  1.422  &  0.559  &  3.266  &  0.566  &  2.385 & \cite{2021Wadhwa} \\
			
			SX Crv  &  0.079  &  0.316638  &  -  &  27.00  &  61.21  &  6340  &  6160  &  1.246  &  0.098  &  1.347  &  0.409  &  2.630  &  0.216  &  2.153 & \cite{2004V402Aur}\\
			
			V53  &  0.079  &  0.308449  &  0.589  &  60.40  &  74.80  &  7415  &  6791  &  1.472  &  0.115  &  1.383  &  0.481  &  5.188  &  0.441  &  2.237 & \cite{2017V53}\\
			
			CRTS J224827.6 + 341351  &  0.079  &  0.321026  &  77.1  &  91.60  &  83.20  &  6077  &  5541  &  1.233  &  0.098  &  1.350  &  0.456  &  2.230  &  0.176  &  2.166 & \cite{2023CRTSLiu}\\
			
			AW UMa  &  0.080  &  0.438700  &  -2.05  &  80.20  &  80.60  &  7175  &  7110  &  1.790  &  0.140  &  1.880  &  0.660  &  8.404  &  0.999  &  3.019 & \cite{2008AWUMa}\\
			
			KIC 10007533  &  0.081  &  0.648064  &  -  &  54.00  &  80.70  &  6808  &  6338  &  2.199  &  0.178  &  2.566  &  0.881  &  12.691  &  1.124  &  4.198 & \cite{2016SenavcKIC}\\
			
			CRTS J155009.2 + 493639  &  0.082  &  0.460910  &  6.22  &  18.90  &  85.70  &  6619  &  6199  &  1.599  &  0.130  &  1.882  &  0.644  &  6.100  &  0.549  &  3.008 & \cite{2023CRTSLiu}\\
			
			V870 Ara  &  0.082  &  0.399722  &  -  &  96.40  &  70.00  &  5860  &  6210  &  1.503  &  0.123  &  1.670  &  0.610  &  2.951  &  0.496  &  2.680 & \cite{2007V870Ara}\\
			
			KIC 8145477  &  0.082  &  0.565784  &  -  &  40.00  &  82.00  &  6801  &  6473  &  2.012  &  0.165  &  2.260  &  0.767  &  9.804  &  0.927  &  3.724 & \cite{2016SenavcKIC}\\
			
			CRTS J234634.7 + 222824  &  0.086  &  0.290693  &  2.19  &  37.30  &  74.60  &  5851  &  5746  &  1.141  &  0.098  &  1.218  &  0.425  &  1.560  &  0.177  &  1.980 & \cite{2023CRTSLiu}\\
			
			TYC 835-1081-1  &  0.086  &  0.448058  &  -  &  66.00  &  80.15  &  6720  &  6480  &  1.470  &  0.130  &  1.760  &  0.630  &  5.668  &  0.628  &  2.870 & \cite{2022TYC835}\\
			
			CRTS J154254.0 + 324652  &  0.087  &  0.354988  &  -8.44  &  94.30  &  84.40  &  5885  &  6067  &  1.317  &  0.114  &  1.464  &  0.514  &  2.307  &  0.321  &  2.373 & \cite{2023CRTSLiu}\\
			
			KIC 11144556  &  0.087  &  0.642980  &  -  &  73.00  &  69.70  &  6428  &  6202  &  2.174  &  0.191  &  2.542  &  0.927  &  9.898  &  1.141  &  4.169& \cite{2016SenavcKIC} \\
			
			KIC 10596883  &  0.088  &  0.468911  &  -  &  18.00  &  69.90  &  7296  &  6514  &  1.772  &  0.156  &  1.882  &  0.641  &  9.005  &  0.664  &  3.155 & \cite{2016SenavcKIC}\\
			
			CRTS J155106.5 + 303534  &  0.089  &  0.380989  &  -5.13  &  69.30  &  74.80  &  7147  &  6819  &  1.384  &  0.122  &  1.559  &  0.552  &  5.690  &  0.591  &  2.530 & \cite{2023CRTSLiu}\\
			
			1SWASP J132829.37+555246.1  &  0.089  &  0.384705  &  -4.46  &  70.00  &  81.50  &  6300  &  6319  &  1.230  &  0.110  &  1.490  &  0.550  &  3.138  &  0.433  &  2.449 & \cite{2021Li}\\
			
			ASAS J103737-3709.5  &  0.090  &  0.343402  &  11.5  &  57.00  &  68.10  &  6050  &  5741  &  1.180  &  0.110  &  1.360  &  0.490  &  2.223  &  0.234  &  2.242 & \cite{2023WadhwaASAS}\\
			
			KIC 8804824  &  0.091  &  0.457404  &  -  &  14.00  &  81.00  &  7202  &  6168  &  1.738  &  0.158  &  1.829  &  0.628  &  8.075  &  0.512  &  3.086 & \cite{2016SenavcKIC} \\
			
			KIC 5374883  &  0.092  &  0.419717  &  -  &  71.66  &  65.92  &  5800  &  5679  &  1.540  &  0.140  &  1.720  &  0.620  &  3.004  &  0.359  &  2.799 & \cite{2020ZhuLiKIC}\\
			
			KR Com  &  0.093  &  0.407968  &  -  &  99.00  &  54.40  &  5920  &  6130  &  0.880  &  0.080  &  1.445  &  0.505  &  2.301  &  0.323  &  2.279 & \cite{2021Gazease-HVAqr}\\
			
			CRTS J170307.9 + 020101  &  0.093  &  0.290884  &  2.54  &  69.80  &  75.70  &  5433  &  5237  &  1.134  &  0.105  &  1.204  &  0.436  &  1.133  &  0.128  &  1.980 & \cite{2023CRTSLiu}\\
			
			CRTS J223837.9 + 321932  &  0.093  &  0.444170  &  5.28  &  44.90  &  74.00  &  6919  &  6756  &  1.541  &  0.144  &  1.784  &  0.646  &  6.544  &  0.780  &  2.910 & \cite{2023CRTSLiu}\\
			
			KIC 7698650  &  0.095  &  0.599155  &  -  &  61.00  &  79.10  &  6107  &  6026  &  2.064  &  0.196  &  2.357  &  0.876  &  6.933  &  0.908  &  3.917 & \cite{2016SenavcKIC}\\
			
			CRTS J164000.2 + 491335  &  0.095  &  0.390782  &  -3.19  &  28.70  &  83.50  &  7137  &  6574  &  1.402  &  0.133  &  1.580  &  0.577  &  5.811  &  0.558  &  2.590 & \cite{2023CRTSLiu}\\
			
			ASAS J153433+1225.3  &  0.096  &  0.330057  &  6.8  &  56.10  &  78.80  &  6079  &  6060  &  1.470  &  0.140  &  1.410  &  0.530  &  2.436  &  0.340  &  2.351 & \cite{2022Li_tensys}\\
			
			CSS J233332.9+180430  &  0.096  &  0.628784  &  -298  &  37.40  &  82.50  &  6770  &  6379  &  2.030  &  0.200  &  2.400  &  0.870  &  10.856  &  1.124  &  4.027 & \cite{2022Li_tensys}\\
			
			FP Boo  &  0.096  &  0.640482  &  -  &  38.00  &  68.80  &  6980  &  6456  &  1.614  &  0.154  &  2.310  &  0.774  &  11.364  &  0.934  &  6.290 & \cite{2006GazeasFPBoo}\\
			
			KIC 11097678  &  0.097  &  0.999716  &  -  &  87.00  &  85.14  &  6493  &  6426  &  0.960  &  0.189  &  3.897  &  1.264  &  24.218  &  2.444  &  4.398 & \cite{2016SenavcKIC}\\
			
			CRTS J162327.1 + 031900  &  0.097  &  0.474562  &  -  &  27.40  &  82.50  &  6914  &  6380  &  1.612  &  0.156  &  1.888  &  0.695  &  7.308  &  0.718  &  3.090 & \cite{2023CRTSLiu}\\
			
			CRTS J133031.1 + 161202  &  0.098  &  0.302666  &  -  &  56.00  &  88.30  &  5860  &  6019  &  1.162  &  0.114  &  1.237  &  0.459  &  1.619  &  0.248  &  2.054 & \cite{2023CRTSLiu}\\
			
			KIC 9453192  &  0.099  &  0.718837  &  -  &  44.00  &  76.90  &  6729  &  6218  &  2.314  &  0.229  &  2.734  &  1.010  &  13.750  &  1.368  &  4.600 & \cite{2016SenavcKIC}\\
			
			NSVS 4701980  &  0.099  &  0.355763  &  -34.4  &  55.30  &  87.40  &  5892  &  6075  &  1.510  &  0.150  &  1.510  &  0.590  &  2.465  &  0.425  &  2.497 & \cite{2022Li_tensys}\\
			
			CRTS J160755.2 + 332342  &  0.099  &  0.357288  &  -  &  36.10  &  74.90  &  7987  &  7476  &  1.310  &  0.130  &  1.445  &  0.537  &  7.624  &  0.808  &  2.388 & \cite{2023CRTSLiu}\\
			
			\hline
	\end{tabular}}
	
\end{table}

\newpage
\begin{table}[h]
	\caption{Absolute parameters of low mass-ratio contact binaries} 
	\resizebox{\textwidth}{!}
	{
		\begin{tabular}{lccccccccccccccl}
			
			\label{tab:abslmr}\\
			\hline
			Star  &  $q$    &  $P$  &   $dP/dt$   &  $f $  &  $i$    &  $T_1$ &  $T_2$ &  $M_1 $  &  $M_2$  &  $R_1$  &   $R_2$   &   $L_1$  &  $L_2$  &  $a$  & Ref.\\
			
			&  ($m_2/m_1$)  & (d) & ($\times10^{-7}$ d)  & (\%)  & $(^0)$ &  (K)  & (K)  & ($M_{\odot}$) &  ($M_{\odot}$) & ($R_{\odot}$)  &  ($R_{\odot}$) & ($L_{\odot}$) & ($L_{\odot}$) & ($R_{\odot}$) & \\
			
			\hline	
			
			NW Aps  &  0.100  &  1.065550  &  -  &  45.00  &  85.10  &  6640  &  6229  &  1.400  &  0.140  &  2.556  &  0.946  &  11.394  
			&  1.209  &  5.060 & \cite{2005NWAPs}\\
			
			AW CrB  &  0.101  &  0.360900  &  3.58  &  75.00  &  82.10  &  6700  &  6808  &  1.390  &  0.140  &  1.480  &  0.580  &  3.960  &  0.648  &  2.453 & \cite{2013AWCrb}\\
			
			DN Boo  &  0.103  &  0.447568  &  -  &  64.00  &  60.02  &  6095  &  6071  &  1.428  &  0.148  &  1.710  &  0.670  &  3.621  &  0.547  &  2.860 & \cite{2008DNBoo}\\
			
			ASAS J082243+1927.0  &  0.106  &  0.280000  &  -  &  72.00  &  76.58  &  5960  &  6078  &  1.100  &  0.170  &  1.150  &  0.420  &  1.497  &  0.216  &  1.947 & \cite{2015J0822}\\
			
			V1191 Cyg  &  0.107  &  0.313400  &  4.5  &  68.60  &  80.40  &  6500  &  6626  &  1.310  &  0.140  &  1.310  &  0.520  &  2.748  &  0.468  &  2.193 & \cite{2012Ulas}\\
			
			CK Boo  &  0.109  &  0.355200  &  0.98  &  65.00  &  64.90  &  6200  &  6291  &  1.430  &  0.160  &  1.440  &  0.560  &  2.749  &  0.441  &  2.459 & \cite{2012CKBoo}\\
			
			NSVS 10368868  &  0.110  &  0.287160  &  10  &  61.00  &  78.00  &  5963  &  6202  &  1.390  &  0.150  &  1.250  &  0.490  &  1.772  &  0.319  &  2.111 & \cite{2022Li_tensys} \\
			
			TYC 6995-813-1  &  0.111  &  0.383180  &  -  &  72.00  &  84.03  &  6300  &  6235  &  1.230  &  0.135  &  1.460  &  0.600  &  3.013  &  0.488  &  2.458  & \cite{2021TYC6995NSVS136}\\
			
			ATO J022.8782+55.2989   &  0.111  &  0.378150  &  -  &  4.50  &  71.75  &  6400  &  6300  &  1.320  &  0.150  &  1.430  &  0.531  &  3.078  &  0.399  &  2.500 & \cite{2018ATOJ1025}\\
			
			FG Hya  &  0.112  &  0.327800  &  -1.96  &  85.60  &  82.30  &  5900  &  6012  &  1.450  &  0.160  &  1.400  &  0.580  &  2.131  &  0.394  &  2.340 & \cite{2005QianFGHya}\\
			
			CSS J110658.4+511201  &  0.119  &  0.400968  &  67  &  92.60  &  86.40  &  6224  &  6291  &  1.560  &  0.180  &  1.650  &  0.700  &  3.666  &  0.689  &  2.747 & \cite{2022Li_tensys}\\
			
			ASASSN-V J022733.60+360447.1  &  0.120  &  0.413965  &  8.5  &  58.30  &  74.60  &  6494  &  6374  &  1.580  &  0.190  &  1.650  &  0.680  &  4.344  &  0.685  &  2.822 & \cite{2022Li_tensys}\\
			
			V2787 Ori  &  0.120  &  0.810980  &  -  &  18.10  &  84.74  &  6993  &  5418  &  1.440  &  0.170  &  2.450  &  0.960  &  12.879  &  0.713  &  4.290 & \cite{2019V2787Ori}\\
			
			V0566 Cam  &  0.121  &  0.444062  &  56.8  &  20.30  &  73.10  &  7037  &  6642  &  1.630  &  0.200  &  1.710  &  0.670  &  6.433  &  0.784  &  2.990& \cite{2022Li_tensys}\\
			
			GR Vir  &  0.122  &  0.346979  &  -4.32  &  78.60  &  83.36  &  6300  &  6163  &  1.370  &  0.170  &  1.420  &  0.610  &  2.850  &  0.482  &  2.395 & \cite{2004QianGRVir}\\
			
			NSVS 6798913  &  0.128  &  0.362541  &  41  &  57.30  &  87.50  &  6029  &  6071  &  1.480  &  0.190  &  1.470  &  0.620  &  2.562  &  0.469  &  2.534& \cite{2022Li_tensys} \\
			
			NSVS 7480723  &  0.128  &  0.299294  &  11.5  &  45.20  &  85.50  &  5822  &  5810  &  1.380  &  0.180  &  1.260  &  0.520  &  1.637  &  0.276  &  2.180 & \cite{2022Li_tensys}\\
			
			eps CrA  &  0.128  &  0.591440  &  4.67  &  25.00  &  74.00  &  6678  &  6341  &  1.700  &  0.230  &  2.100  &  0.850  &  7.869  &  1.048  &  3.690 & \cite{2005epsCrA}\\
			
			
			ASAS0154+20  &  0.137  &  0.472488  &  -  &  16.00  &  79.90  &  6680  &  6011  &  1.916  &  0.263  &  1.858  &  0.768  &  6.167  &  0.691  &  3.303 & \cite{2022J0154}\\
			
			V584 Cam  &  0.137  &  0.464699  &  -  &  33.00  &  68.42  &  5573  &  5321  &  1.546  &  0.212  &  1.754  &  0.728  &  2.663  &  0.381  &  3.041 & \cite{2021V584_NX_NSVS}\\
			
			
			V345 Gem  &  0.142  &  0.274800  &  0.59  &  73.30  &  72.90  &  6115  &  6365  &  1.060  &  0.150  &  1.100  &  0.500  &  1.518  &  0.368  &  1.892 & \cite{2009V345Gem}\\
			
			NSVS 3650324  &  0.142  &  0.379297  &  50.6  &  96.30  &  77.80  &  6590  &  6463  &  1.490  &  0.210  &  1.550  &  0.720  &  4.065  &  0.812  &  2.627 & \cite{2022Li_tensys}\\
			
			V410 Aur  &  0.143  &  0.366400  &  8.22  &  52.40  &  78.60  &  6040  &  5915  &  1.300  &  0.190  &  1.400  &  0.610  &  2.340  &  0.409  &  2.456 & \cite{2005V410Aur}\\
			
			V710 Mon  &  0.143  &  0.405200  &  1.95  &  62.70  &  79.90  &  6145  &  6294  &  1.140  &  0.160  &  1.460  &  0.660  &  2.727  &  0.613  &  2.512 & \cite{2014V710Mon}\\
			
			ASAS J142124+1813.1  &  0.144  &  0.242720  &  -  &  23.00  &  79.79  &  6160  &  6970  &  3.144  &  0.453  &  1.403  &  0.597  &  2.543  &  0.755  &  2.510 & \cite{2019ASASJ142}\\
			
			
			TIC 89428764  &  0.147  &  0.376610  &  6.36  &  32.70  &  76.90  &  7070  &  6797  &  1.500  &  0.220  &  1.470  &  0.630  &  4.844  &  0.760  &  2.630 & \cite{2022TIC3939-8942}\\
			
			XY LMi  &  0.148  &  0.436900  &  -1.67  &  74.10  &  81.00  &  6144  &  6093  &  1.180  &  0.180  &  1.540  &  0.710  &  3.032  &  0.623  &  2.679 & \cite{2011XYLMi}\\
			
			EM Psc  &  0.149  &  0.343959  &  39.7  &  95.30  &  88.60  &  5300  &  4987  &  1.050  &  0.156  &  1.283  &  0.631  &  1.165  &  0.221  &  2.195 & \cite{2008QianEMPis}\\
			
			NX Cam  &  0.150  &  0.605771  &  -  &  54.60  &  82.36  &  6657  &  6015  &  1.750  &  0.262  &  2.025  &  0.839  &  7.225  &  0.827  &  3.796 & \cite{2021V584_NX_NSVS}\\
			
			NSVS 5029961  &  0.151  &  0.376666  &  -  &  33.80  &  77.30  &  6260  &  6206  &  1.872  &  0.284  &  1.573  &  0.680  &  3.409  &  0.615  &  2.830 & \cite{2021NSVS5029}\\
			
			TIC 321576458  &  0.153  &  0.376145  &  -  &  41.70  &  82.57  &  6140  &  6490  &  1.200  &  0.183  &  1.417  &  0.632  &  2.560  &  0.636  &  2.438 & This study\\
			
			MW Pav  &  0.154  &  0.794997  &  -  &  71.57  &  80.20  &  6900  &  6987  &  1.670  &  0.257  &  2.481  &  1.193  &  12.518  &  3.043  &  4.485 & This study\\
			
			NSVS 2643686  &  0.158  &  0.396068  &  -  &  72.90  &  81.31  &  5905  &  5840  &  1.309  &  0.206  &  1.536  &  0.721  &  2.574  &  0.543  &  2.602 & \cite{2021V584_NX_NSVS}\\
			
			HV Aqr  &  0.159  &  0.374455  &  -1.29  &  56.21  &  78.22  &  5915  &  5948  &  1.050  &  0.160  &  1.320  &  0.580  &  1.914  &  0.378  &  2.325 &  \cite{2023Zubairi} \\
			
			V728 Her  &  0.160  &  0.471300  &  1.92  &  81.00  &  68.70  &  6585  &  6678  &  1.800  &  0.280  &  1.870  &  0.820  &  5.899  &  1.200  &  3.247 & \cite{2016V728Her}\\
			
			CRTS J163819.6+03485  &  0.160  &  0.205332  &  -0.711  &  63.00  &  88.50  &  6696  &  6502  &  1.190  &  0.190  &  0.930  &  0.440  &  1.560  &  0.310  &  1.628 & \cite{2023CRTSJ1638}\\
			
			NSVS 1926064  &  0.160  &  0.407472  &  -  &  70.90  &  78.40  &  6020  &  5923  &  1.558  &  0.249  &  1.605  &  0.755  &  3.035  &  0.629  &  2.812 & \cite{2020NSVS19260}\\
			
			V144  &  0.160  &  0.721590  &  -  &  19.50  &  76.57  &  7120  &  6808  &  1.310  &  0.210  &  2.230  &  0.950  &  11.466  &  1.739  &  3.890 & \cite{2013V144}\\
			
			NSVS 2256852  &  0.161  &  0.348890  &  -2.36  &  17.30  &  75.60  &  6804  &  5606  &  0.950  &  0.150  &  1.180  &  0.520  &  2.677  &  0.240  &  2.160  & \cite{2018NSVS22568}\\
			
			TIC 393943031  &  0.163  &  0.449793  &  4.21  &  50.80  &  81.60  &  6460  &  6270  &  1.400  &  0.230  &  1.620  &  0.750  &  4.101  &  0.780  &  2.910 & \cite{2022TIC3939-8942}\\
			
			AH Aur  &  0.165  &  0.494109  &  -2.75  &  75.00  &  76.10  &  6200  &  6418  &  1.674  &  0.283  &  1.897  &  0.837  &  4.771  &  1.066  &  3.284 & \cite{2005AcA....55..123G}\\
			
			TV Mus  &  0.166  &  0.445675  &  -2.16  &  74.30  &  77.15  &  5980  &  5808  &  1.350  &  0.220  &  1.700  &  0.830  &  3.316  &  0.703  &  2.848 & \cite{2005QianTV-CU}\\
			
			AH Cnc  &  0.168  &  0.360440  &  3.99  &  58.50  &  90.00  &  6300  &  6265  &  1.100  &  0.190  &  1.299  &  0.619  &  2.385  &  0.530  &  2.316 & \cite{2005ZhangAHCnc,2006QianCnc}\\
			
			NSVS 13602901  &  0.171  &  0.523790  &  -  &  44.00  &  83.59  &  6250  &  6222  &  1.190  &  0.203  &  1.690  &  0.790  &  3.910  &  0.839  &  3.048 & \cite{2021TYC6995NSVS136}\\
			
			
			CU Tau  &  0.177  &  0.412537  &  -18.1  &  50.10  &  73.95  &  5900  &  5938  &  1.200  &  0.210  &  1.445  &  0.695  &  2.270  &  0.539  &  2.610 & \cite{2005QianTV-CU}\\
			
			V458 Mon  &  0.183  &  0.396952  &  3.69  &  29.50  &  82.68  &  6144  &  6067  &  1.200  &  0.219  &  1.379  &  0.655  &  2.431  &  0.522  &  2.549 & \cite{2023V458Mon}\\
			
			TY Pup  &  0.184  &  0.819240  &  0.557  &  84.30  &  83.63  &  6900  &  6915  &  1.650  &  0.303  &  2.636  &  1.373  &  14.131  &  3.867  &  4.597 & \cite{2018TYPuP}\\
			GV Leo  &  0.188  &  0.266730  &  -4.95  &  17.74  &  76.13  &  4850  &  5344  &  1.010  &  0.190  &  1.010  &  0.490  &  0.506  &  0.176  &  1.850 & \cite{2013Kriwattanawong-GVLeo}\\
			
			BO Ari  &  0.190  &  0.318190  &  -2.81  &  49.80  &  81.91  &  5940  &  5908  &  0.995  &  0.189  &  1.090  &  0.515  &  1.327  &  0.290  &  2.071 & \cite{2015BOAri}\\
			
			EK Aqr  &  0.192  &  0.612802  &  3.63  &  32.80  &  76.38  &  7900  &  6808  &  1.795  &  0.345  &  2.113  &  1.030  &  15.603  &  2.045  &  3.905 & \cite{2022EKAqr}\\
			
			Y Sex  &  0.195  &  0.419810  &  0.46  &  50.80  &  77.67  &  6400  &  6211  &  1.559  &  0.305  &  1.589  &  0.799  &  3.801  &  0.852  &  2.898 & \cite{2021YSex_V1363Ori}\\
			
			CSS J022914.4+044340  &  0.198  &  0.306139  &  -  &  63.70  &  88.40  &  5855  &  5747  &  1.440  &  0.290  &  1.260  &  0.650  &  1.674  &  0.414  &  2.290   &  \cite{2021CSS_J0229}\\
			
			V402 Aur  &  0.201  &  0.603500  &  -  &  3.00  &  52.65  &  6700  &  6775  &  1.638  &  0.327  &  0.915  &  1.997  &  1.514  &  7.538  &  3.770 &   \cite{2004V402Aur}\\
			
			V1363 Ori  &  0.205  &  0.431924  &  1.56  &  56.60  &  56.90  &  6156  &  5627  &  1.181  &  0.242  &  1.470  &  0.736  &  2.784  &  0.487  &  2.699 & \cite{2021YSex_V1363Ori}\\
			
			DN Aur  &  0.205  &  0.616889  &  -  &  46.10  &  76.88  &  6830  &  6750  &  1.440  &  0.295  &  1.983  &  1.016  &  7.677  &  1.923  &  3.657  &  \cite{1996goderyaDNAur}\\
			
			
			HI Pup  &  0.206  &  0.432620  &  -  &  20.00  &  82.20  &  6500  &  6377  &  1.220  &  0.230  &  1.440  &  0.670  &  3.321  &  0.666  &  2.720 &   \cite{2014HIPup}\\
			
			V429 Cam  &  0.208  &  0.441160  &  6.97  &  37.90  &  81.50  &  6366  &  6287  &  1.360  &  0.280  &  1.550  &  0.780  &  3.540  &  0.853  &  2.870  & \cite{Li2021AJ}\\
			
			TIC 159102550  &  0.211  &  0.488239  &  -  &  20.74  &  77.21  &  6109  &  6015  &  1.400  &  0.295  &  1.633  &  0.817  &  3.332  &  0.784  &  3.105 &  This study\\
			
			NS Cam  &  0.213  &  0.907390  &  -2.25  &  17.00  &  85.00  &  6250  &  5689  &  1.180  &  0.250  &  2.360  &  1.160  &  7.625  &  1.265  &  4.450 & \cite{2020NSCam}\\
			
			V409 Hya  &  0.216  &  0.472270  &  5.41  &  60.60  &  89.46  &  7000  &  6730  &  1.500  &  0.330  &  1.696  &  0.903  &  6.196  &  1.501  &  3.116  & \cite{2014V409Hya}\\
			
			
			V830 Cep  &  0.228  &  0.260133  &  -0.85  &  26.20  &  88.60  &  6021  &  6147  &  0.840  &  0.190  &  0.910  &  0.470  &  0.976  &  0.283  &  1.729 & \cite{2021Li}\\
			
			YY Crb  &  0.232  &  0.376564  &  -  &  23.00  &  79.50  &  6100  &  6499  &  1.393  &  0.339  &  1.385  &  0.692  &  2.383  &  0.766  &  2.630& \cite{2005AcA....55..123G}\\
			
			KN Per  &  0.236  &  0.866464  &  5.12  &  53.40  &  83.01  &  7200  &  6942  &  2.000  &  0.472  &  2.749  &  1.527  &  18.221  &  4.857  &  5.162 & \cite{1997goderyaKNPer,2022KNPer_Gao}\\
			
			GSC 3208-1986  &  0.237  &  0.404562  &  -  &  39.00  &  85.80  &  6875  &  6760  &  1.440  &  0.341  &  1.469  &  0.794  &  4.325  &  1.181  &  2.785 & \cite{2015GSC3208}\\
			
			V921 Her  &  0.244  &  0.877380  &  -  &  23.00  &  68.10  &  7780  &  7346  &  2.068  &  0.505  &  2.752  &  1.407  &  24.895  &  5.172  &  5.290 & \cite{2006GazeasFPBoo}\\
			
			V1068 Her  &  0.266  &  0.394305  &  1.247  &  14.11  &  83.53  &  6158  &  6143  &  1.050  &  0.279  &  1.252  &  0.696  &  2.022  &  0.618  &  2.482 & This study\\
			\hline
	\end{tabular}}
	
\end{table}

\begin{figure}[h]
	\centering
	\includegraphics[width=0.7\columnwidth]{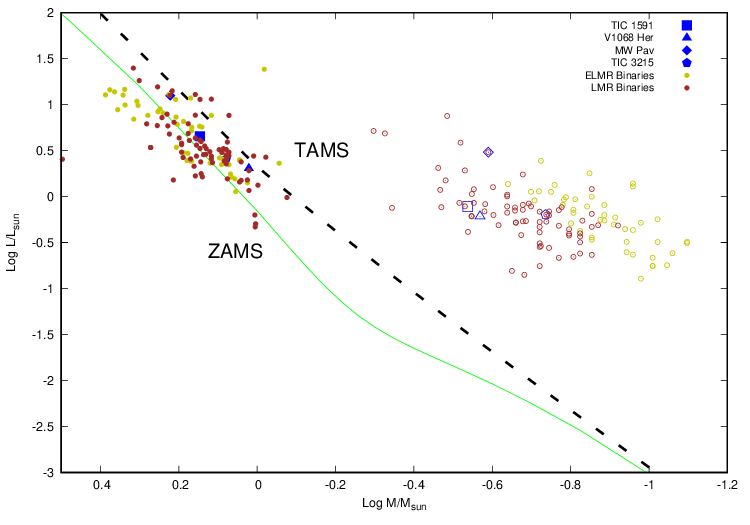}
	\caption{\centering The mass-luminosity diagram of low mass ratio contact binaries. The solid symbols represents the primary components and the hollow are secondary components. Symbols with blue color shows our targets.}
	\label{fig:ML-dia}
\end{figure}

The position of primary and secondary components of our targets along with other binaries listed in Table \ref{tab:abslmr} is shown in Figure \ref{fig:ML-dia}. The filled symbols are chosen to represent primary and hollow symbols represents secondary components of each contact binary system. Our targets are presented in blue color. From Figure \ref{fig:ML-dia}, it is clear that most of the primary components, along with our observed targets, are residing around Zero-age Main-Sequence (ZAMS), whereas, the less massive secondaries are above the Terminal Age Main-Sequence (TAMS) line. This indicates that primary stars are little or un-evolved and secondaries have evolved out of main sequence zone. Our four targets are also following the same trend.

According to \cite{2013YildizOriginofWUMa, 2014Yildiz_massAge} the initial mass of both, primary and secondary, components also help us understand the evolutionary process of contact binaries. They proposed a new method and calculate the initial mass of primary and secondary components of a binary system. They also estimated the age of contact binary systems. We have employed the model proposed by \cite{2013YildizOriginofWUMa} to calculate the initial mass of both components. We have computed the initial mass of secondary component ($M_{2i}$) in terms of its higher luminosity for present mass using Equation \ref{eq:m2i}, with $M_2$ as the present mass of secondary and $M_{L2}$ as $( L_2 / 1.49 )^{1/4.216}$ in solar units. The initial mass of the primary component $M_{1i}$ is calculated by using $\gamma=0.6640$, the same value as used by \cite{2013YildizOriginofWUMa}. In order to calculate the age of our sample stars we have used the following equations \ref{eq:M2mean} and \ref{eq:M2age}. We have also compared our target with other low mass ratio binaries. The values of their initial masses are tabulated in Table \ref{tab:initialMass}

\begin{equation}
	\centering
	M_{2i} = M_2 + 2.50\times(M_{L2} - M_2 -0.07)^{0.64} ,
	\label{eq:m2i}
\end{equation}

\begin{equation}
	\centering
	M_{2mean} = \frac{M_{2i}+M_{L2}}{2}
	\label{eq:M2mean}
\end{equation}

\begin{equation}
	\centering
	\tau_{MS} = \frac{10}{(M/M_{\odot})^{4.05}} \times \left[0.00560 \times \left(\frac{M}{M_{\odot}}+3.993\right)^{3.16} + 0.042 \right] Gyr
	\label{eq:M2age}
\end{equation}

For these binaries the ratio of spin angular momentum ($J_s$) and orbital angular momentum ($J_o$) is calculated using the equation \ref{eq:jojs} provided by \cite{2015Yang} and is listed in Table \ref{tab:initialMass}

\begin{equation}
	\frac{J_s}{J_o} = \frac{1+q}{q} \left( k^2_1r^2_1 + qk^2_2r^2_2  \right)\, ,
	\label{eq:jojs}
\end{equation}

The value of gyration radii ($k^2_1 = k^2_2=k^2$) was opted from \citep{2006Li_gyration} as $0.06$. The ratio of spin and orbital angular momentum indicates that all LMR, including our targets, are in dynamically stable state. However, a few of the ELMR binaries have crossed the threshold value of $1/3$. According to \cite{2021Li}, the most likely reason of this, is that the gyration radii (k) must be lower than 0.06. This means that the limit of mass ratio of contact binaries is affected by $k^2$. 

\begin{figure}[h]
	\centering
	\includegraphics[width=0.49\columnwidth]{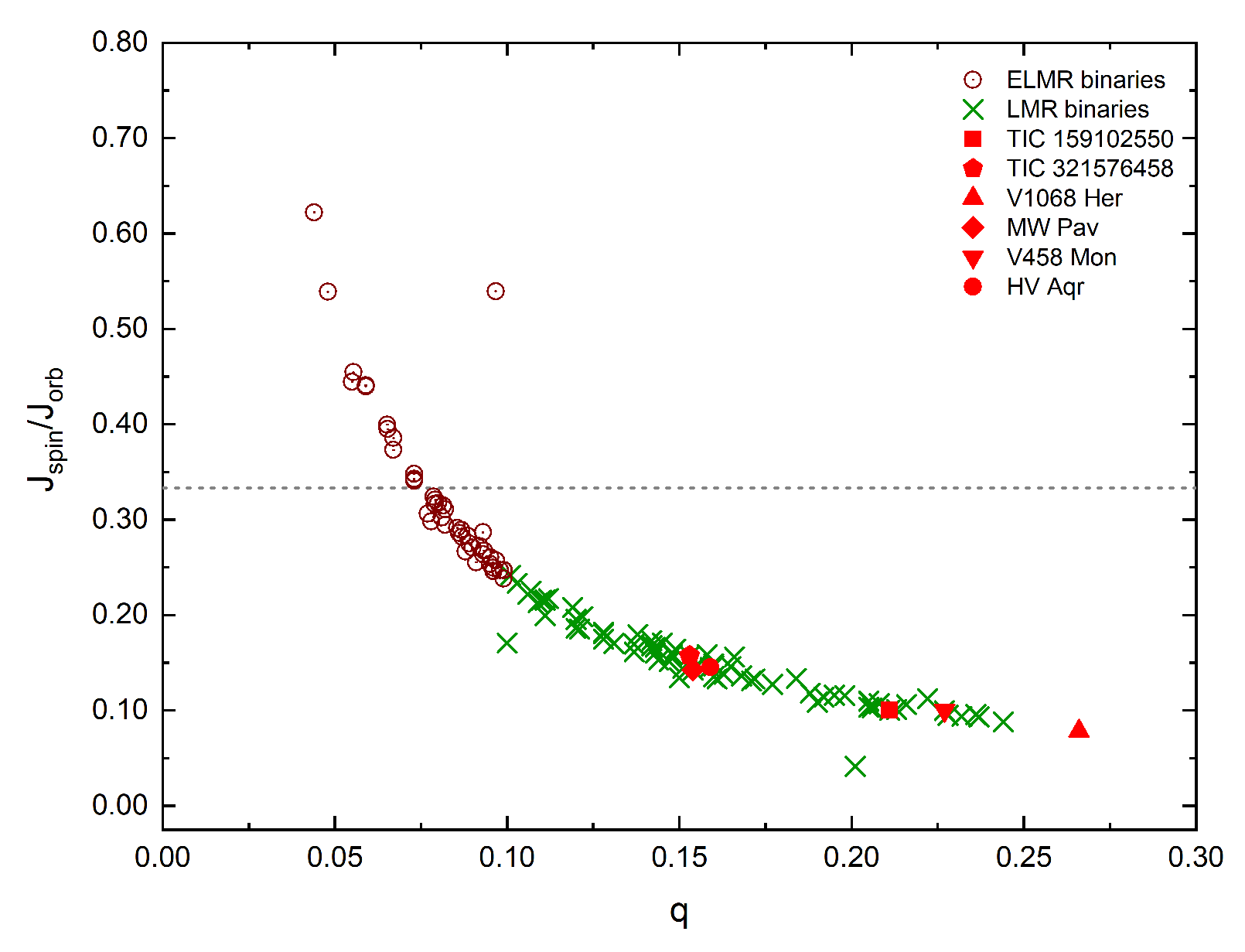}
	\includegraphics[width=0.49\columnwidth]{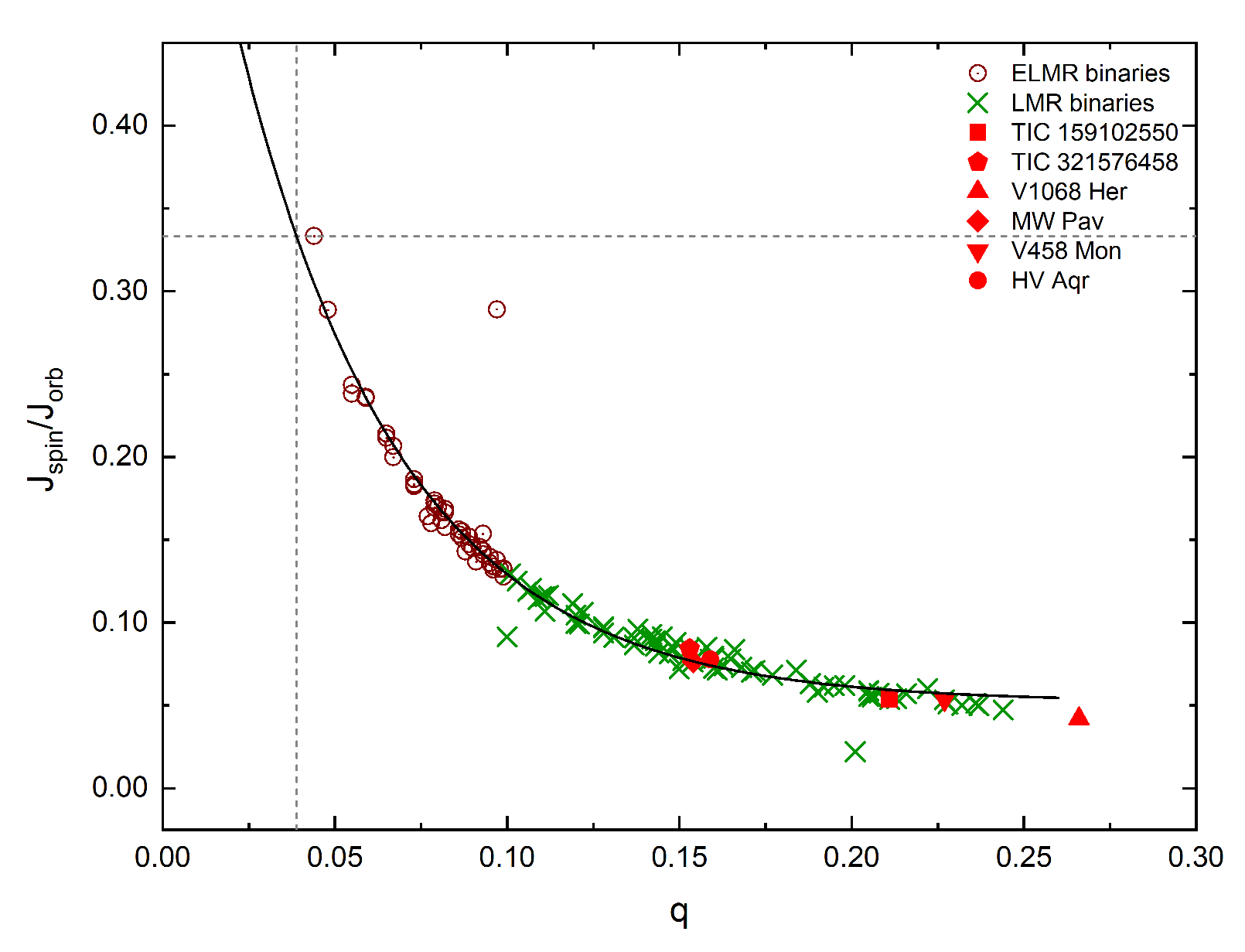}
	\caption{\centering The relationship with mass ratio and $J_s/J_o$. The horizontal dashed line represents critical value for Darwin's stability. Left Figure shows the values with $k^2=0.06$ and right one with $k^2 = 0.03214$}
	\label{fig:jojs-dia}

\end{figure}

If we keep the system V1187 Her exactly at the stability limit, then the value of $k^2$ is 0.03214. The relation between q and spin-orbital angular momentum ratio is shown in figure \ref{fig:jojs-dia}. The fitting function is shown in Equation \ref{eq:mew-jojs}.

\begin{equation}
	\frac{J_s}{J_o} = 0.052(\pm0.002) + 0.64(\pm0.02) \times e^{-21.15(\pm0.48)\times q} 
	\label{eq:mew-jojs}
\end{equation}

Based on our samples, the updated relationship between mass ratio and spin-orbital angular momentum ratio gives us a new value of $q_{min} = 0.0388$ for Darwin's stability in contact binaries. 

\section{Acknowledgments}
This study is partly supported by the National Natural Science Foundation of China (Grant Nos. 11933008), Yunnan Natural Science Foundation (No. 202001AT070051) and Yunnan Revitalization Talent Support Program Young Talent Project. Some data of Present study is from LAMOST, which is operated and managed by the National Astronomical Observatories, Chinese Academy of Sciences. This work also used TESS and GAIA data and authors thanks a lot for the team of TESS and GAIA. This research has also made use of the International Variable Star Index (VSX) database, operated at AAVSO, Cambridge, Massachusetts, USA.

\newpage
\begin{longtable}[h]{lcccccc}
	
	\caption{\centering Momentum ratio,  and initial mass of primary and secondary components of selected binary systems}	 
	\label{tab:initialMass}\\
	\hline
	Star  &  q  &  $J_{spin} / J_{orb}^*$  &  $J_{spin} / J_{orb}$  &  $M_{2i}$  &  $M_{1i}$ &  Age \\
	\endfirsthead
	
	\multicolumn{6}{c}{Continuation of Table \ref{tab:initialMass}}\\
	\hline
	
	Star  &  q & $J_{spin} / J_{orb}^*$  &  $J_{spin} / J_{orb}$  &  $M_{2i}$  &  $M_{1i}$ &  Age \\
	
	\hline
	\endhead
	
	\hline
	\endfoot
	
	\hline

	V1187 Her   &   0.044   &   0.3309   &   0.6147   &   1.737   &   0.633   &   6.63     \\
	TYC 4002-2628-1   &   0.048   &   0.2901   &   0.5389   &   1.586   &   0.595   &   9.12      \\
	VSX J082700.8+462850   &   0.055   &   0.2393   &   0.4446   &   1.471   &   0.586   &   11.88      \\
	IP Lyn   &   0.055   &   0.2448   &   0.4546   &   1.918   &   1.294   &   4.58      \\
	KIC 4244929   &   0.059   &   0.2375   &   0.4411   &   1.676   &   0.947   &   7.35      \\
	KIC 9151972   &   0.059   &   0.2367   &   0.4398   &   1.757   &   1.048   &   6.22      \\
	ASAS J083241+2332.4   &   0.065   &   0.2151   &   0.3996   &   1.694   &   0.678   &   7.12      \\
	V857 Her   &   0.065   &   0.2126   &   0.3949   &   2.365   &   1.433   &   2.23      \\
	KIC 8539720   &   0.067   &   0.2008   &   0.3731   &   2.102   &   1.786   &   3.28      \\
	CW Lyn   &   0.067   &   0.2076   &   0.3857   &   2.254   &   0.960   &   2.64     \\
	ASAS J104422-0711.2   &   0.073   &   0.1874   &   0.3481   &   2.282   &   0.750   &   2.54     \\
	KIC 3127873   &   0.073   &   0.1845   &   0.3427   &   2.026   &   1.643   &   3.70     \\
	KIC 12352712   &   0.073   &   0.1833   &   0.3406   &   2.219   &   1.690   &   2.73     \\
	NSVS 2569022   &   0.077   &   0.1649   &   0.3063   &   1.563   &   0.675   &   9.38     \\
	ZZ PsA   &   0.078   &   0.1607   &   0.2984   &   1.993   &   0.575   &   4.03     \\
	SX Crv   &   0.079   &   0.1745   &   0.3241   &   1.624   &   0.733   &   8.16     \\
	V53   &   0.079   &   0.1703   &   0.3163   &   1.867   &   0.883   &   5.00     \\
	CRTS J224827.6 + 341351   &   0.079   &   0.1727   &   0.3208   &   1.726   &   0.686   &   6.61    \\
	AW UMa   &   0.080   &   0.1707   &   0.3172   &   2.097   &   1.133   &   3.33     \\
	KIC 10007533   &   0.081   &   0.1626   &   0.3021   &   2.145   &   1.538   &   3.05     \\
	CRTS J155009.2 + 493639   &   0.082   &   0.1694   &   0.3147   &   2.021   &   0.964   &   3.79     \\
	V870 Ara   &   0.082   &   0.1673   &   0.3108   &   1.885   &   0.911   &   4.82     \\
	KIC 8145477   &   0.082   &   0.1585   &   0.2944   &   2.076   &   1.370   &   3.42     \\
	CRTS J234634.7 + 222824   &   0.086   &   0.1569   &   0.2914   &   1.652   &   0.619   &   7.68     \\
	TYC 835-1081-1   &   0.086   &   0.1540   &   0.2860   &   1.961   &   0.855   &   4.20     \\
	CRTS J154254.0 + 324652   &   0.087   &   0.1557   &   0.2893   &   1.806   &   0.749   &   5.60     \\
	KIC 11144556   &   0.087   &   0.1518   &   0.2820   &   2.136   &   1.521   &   3.08     \\
	KIC 10596883   &   0.088   &   0.1435   &   0.2666   &   1.957   &   1.167   &   4.18     \\
	CRTS J155106.5 + 303534   &   0.089   &   0.1525   &   0.2833   &   1.862   &   0.799   &   5.02     \\
	1SWASP J132829.37+555246.1   &   0.089   &   0.1480   &   0.2750   &   1.844   &   0.647   &   5.22     \\
	ASAS J103737-3709.5   &   0.090   &   0.1456   &   0.2705   &   1.641   &   0.666   &   7.82     \\
	KIC 8804824   &   0.091   &   0.1375   &   0.2554   &   1.859   &   1.167   &   4.97     \\
	KIC 5374883   &   0.092   &   0.1465   &   0.2721   &   1.752   &   0.998   &   6.13     \\
	KR Com   &   0.093   &   0.1543   &   0.2867   &   1.778   &   0.309   &   6.01     \\
	CRTS J170307.9 + 020101   &   0.093   &   0.1420   &   0.2638   &   1.694   &   0.600   &   7.02     \\
	CRTS J223837.9 + 321932   &   0.093   &   0.1440   &   0.2676   &   1.991   &   0.920   &   3.96     \\
	KIC 7698650   &   0.095   &   0.1365   &   0.2537   &   2.042   &   1.444   &   3.57      \\
	CRTS J164000.2 + 491335   &   0.095   &   0.1402   &   0.2604   &   1.884   &   0.814   &   4.80     \\
	ASAS J153433+1225.3   &   0.096   &   0.1344   &   0.2497   &   1.722   &   0.938   &   6.50     \\
	CSS J233332.9+180430   &   0.096   &   0.1326   &   0.2463   &   2.128   &   1.382   &   3.10     \\
	FP Boo   &   0.096   &   0.1325   &   0.2461   &   2.085   &   0.965   &   3.38     \\
	KIC 11097678   &   0.097   &   0.2905   &   0.5396   &   2.439   &   0.204   &   1.99     \\
	CRTS J162327.1 + 031900   &   0.097   &   0.1384   &   0.2571   &   2.063   &   0.971   &   3.50     \\
	CRTS J133031.1 + 161202   &   0.098   &   0.1331   &   0.2472   &   1.684   &   0.634   &   7.13     \\
	KIC 9453192   &   0.099   &   0.1283   &   0.2384   &   2.183   &   1.657   &   2.83     \\
	NSVS 4701980   &   0.099   &   0.1331   &   0.2473   &   1.788   &   0.960   &   5.70     \\
	CRTS J160755.2 + 332342   &   0.099   &   0.1329   &   0.2470   &   1.811   &   0.745   &   5.50     \\
	
	NW Aps   &   0.100   &   0.0919   &   0.1707   &   2.205   &   0.706   &   2.82     \\
	AW CrB   &   0.101   &   0.1301   &   0.2417   &   1.965   &   0.777   &   4.15     \\
	DN Boo   &   0.103   &   0.1256   &   0.2334   &   1.902   &   0.839   &   4.62     \\
	ASAS J082243+1927.0   &   0.106   &   0.1193   &   0.2216   &   1.550   &   0.636   &   9.16     \\
	V1191 Cyg   &   0.107   &   0.1212   &   0.2252   &   1.831   &   0.742   &   5.28     \\
	CK Boo   &   0.109   &   0.1146   &   0.2128   &   1.794   &   0.881   &   5.60     \\
	NSVS 10368868   &   0.110   &   0.1162   &   0.2158   &   1.701   &   0.869   &   6.75     \\
	TYC 6995-813-1   &   0.111   &   0.1162   &   0.2159   &   2.013   &   0.599   &   3.83     \\
	ATO J022.8782+55.2989    &   0.111   &   0.1074   &   0.1995   &   1.770   &   0.776   &   5.89     \\
	FG Hya   &   0.112   &   0.1170   &   0.2173   &   1.759   &   0.913   &   5.98     \\
	CSS J110658.4+511201   &   0.119   &   0.1119   &   0.2079   &   1.933   &   0.971   &   4.31     \\
	ASASSN-V J022733.60+360447.1   &   0.120   &   0.1052   &   0.1953   &   1.935   &   0.994   &   4.28     \\
	V2787 Ori   &   0.120   &   0.1001   &   0.1860   &   1.970   &   0.835   &   4.06     \\
	V0566 Cam   &   0.121   &   0.0997   &   0.1852   &   1.984   &   1.031   &   3.92     \\
	GR Vir   &   0.122   &   0.1068   &   0.1983   &   1.824   &   0.814   &   5.27     \\
	NSVS 6798913   &   0.128   &   0.0980   &   0.1820   &   1.795   &   0.941   &   5.50     \\
	NSVS 7480723   &   0.128   &   0.0972   &   0.1806   &   1.621   &   0.896   &   7.82     \\
	eps CrA   &   0.128   &   0.0941   &   0.1748   &   2.060   &   1.085   &   3.42     \\
	ASAS J102556+2049.3   &   0.131   &   0.0916   &   0.1702   &   1.601   &   0.619   &   8.36     \\
	ASAS0154+20   &   0.137   &   0.0868   &   0.1612   &   1.870   &   1.376   &   4.64     \\
	V584 Cam   &   0.137   &   0.0912   &   0.1694   &   1.762   &   1.025   &   5.79     \\
	V776 Cas   &   0.138   &   0.0967   &   0.1797   &   2.047   &   1.079   &   3.49     \\
	V345 Gem   &   0.142   &   0.0904   &   0.1679   &   1.742   &   0.525   &   6.22     \\
	NSVS 3650324   &   0.142   &   0.0932   &   0.1732   &   1.980   &   0.895   &   3.93     \\
	V410 Aur   &   0.143   &   0.0861   &   0.1600   &   1.745   &   0.777   &   6.05     \\
	V710 Mon   &   0.143   &   0.0896   &   0.1664   &   1.924   &   0.547   &   4.42     \\
	DZ Psc   &   0.145   &   0.0918   &   0.1706   &   1.820   &   0.802   &   5.27     \\
	TIC 89428764   &   0.147   &   0.0815   &   0.1514   &   1.950   &   0.919   &   4.12     \\
	XY LMi   &   0.148   &   0.0854   &   0.1586   &   1.909   &   0.599   &   4.49     \\
	EM Psc   &   0.149   &   0.0883   &   0.1641   &   1.569   &   0.575   &   8.88     \\
	NX Cam   &   0.150   &   0.0723   &   0.1343   &   1.942   &   1.186   &   4.10     \\
	NSVS 5029961   &   0.151   &   0.0774   &   0.1438   &   1.795   &   1.364   &   5.26    \\
	TIC 321576458   &   0.153   &   0.0847   &   0.1573   &   1.917   &   0.617   &   4.43     \\
	MW Pav   &   0.154   &   0.0767   &   0.1425   &   2.523   &   0.909   &   1.76     \\
	NSVS 2643686   &   0.158   &   0.0854   &   0.1587   &   1.832   &   0.763   &   5.10     \\
	HV Aqr   &   0.159   &   0.0782   &   0.1453   &   1.777   &   0.507   &   5.78     \\
	V728 Her   &   0.160   &   0.0801   &   0.1487   &   2.083   &   1.194   &   3.24     \\
	CRTS J163819.6+03485   &   0.160   &   0.0792   &   0.1471   &   1.645   &   0.701   &   7.40     \\
	NSVS 1926064   &   0.160   &   0.0788   &   0.1463   &   1.852   &   1.019   &   4.82    \\
	V144   &   0.160   &   0.0729   &   0.1354   &   2.290   &   0.611   &   2.43     \\
	NSVS 2256852   &   0.161   &   0.0717   &   0.1332   &   1.603   &   0.462   &   8.27     \\
	TIC 393943031   &   0.163   &   0.0741   &   0.1377   &   1.950   &   0.822   &   4.10     \\
	AH Aur   &   0.165   &   0.0786   &   0.1459   &   2.038   &   1.084   &   3.47     \\
	TV Mus   &   0.166   &   0.0841   &   0.1562   &   1.923   &   0.778   &   4.31     \\
	AH Cnc   &   0.168   &   0.0734   &   0.1363   &   1.512   &   0.656   &   9.83     \\
	NSVS 13602901   &   0.171   &   0.0705   &   0.1310   &   1.861   &   0.633   &   4.85      \\
	AS CrB   &   0.172   &   0.0716   &   0.1330   &   1.938   &   0.669   &   4.22     \\
	CU Tau   &   0.177   &   0.0685   &   0.1273   &   1.569   &   0.743   &   8.58     \\
	V458 Mon   &   0.183   &   0.0636   &   0.1182   &   1.805   &   0.667   &   5.33     \\
	TY Pup   &   0.184   &   0.0718   &   0.1334   &   2.607   &   0.876   &   1.57     \\
	GV Leo   &   0.188   &   0.0635   &   0.1180   &   1.457   &   0.584   &   11.17     \\
	BO Ari   &   0.190   &   0.0584   &   0.1085   &   1.622   &   0.513   &   7.75     \\
	EK Aqr   &   0.192   &   0.0614   &   0.1140   &   2.266   &   1.150   &   2.41     \\
	Y Sex   &   0.195   &   0.0623   &   0.1158   &   1.910   &   1.020   &   4.25     \\
	CSS J022914.4+044340   &   0.198   &   0.0623   &   0.1157   &   1.633   &   0.989   &   7.14     \\
	V402 Aur   &   0.201   &   0.0223   &   0.0413   &   2.136   &   1.030   &   2.94     \\
	V1363 Ori   &   0.205   &   0.0592   &   0.1100   &   1.752   &   0.674   &   5.81     \\
	DN Aur   &   0.205   &   0.0588   &   0.1093   &   1.756   &   0.949   &   5.61     \\
	HI Pup   &   0.206   &   0.0554   &   0.1028   &   1.907   &   0.656   &   4.41     \\
	V429 Cam   &   0.208   &   0.0576   &   0.1070   &   1.936   &   0.804   &   4.11     \\
	TIC 159102550   &   0.211   &   0.0540   &   0.1003   &   1.886   &   0.865   &   4.45     \\
	NS Cam   &   0.213   &   0.0544   &   0.1010   &   2.134   &   0.547   &   3.02      \\
	V409 Hya   &   0.216   &   0.0573   &   0.1064   &   1.615   &   1.068   &   7.23     \\
	V830 Cep   &   0.228   &   0.0512   &   0.0950   &   1.621   &   0.359   &   7.77     \\
	YY Crb   &   0.232   &   0.0503   &   0.0935   &   1.822   &   0.895   &   4.88      \\
	KN Per   &   0.236   &   0.0515   &   0.0956   &   2.607   &   1.283   &   1.51      \\
	GSC 3208-1986   &   0.237   &   0.0502   &   0.0932   &   1.493   &   1.053   &   9.26     \\
	V921 Her   &   0.244   &   0.0474   &   0.0881   &   2.609   &   1.361   &   1.49      \\
	V1068 Her   &   0.266   &   0.0423   &   0.0786   &   1.805   &   0.537   &   5.17     \\
	\hline
	
\end{longtable}
\vspace{-0.4cm}

\hspace{2.1cm}  \small{*Revised value}

 \bibliography{ref}{}
 \bibliographystyle{aasjournal}

 \end{document}